\documentclass[twocolumn]{article}
\usepackage{times,bm}
\usepackage{amsmath}
\usepackage[affil-it,]{authblk}
\usepackage[pagebackref=false,colorlinks=true,linkcolor=acsblue,citecolor=olive,urlcolor=acsblue]{hyperref}

\usepackage{geometry}
\usepackage[switch*, displaymath,pagewise, mathlines]{lineno}
\usepackage{graphicx}
\usepackage{cite}
\usepackage{amsmath}
\usepackage{amsfonts}
\usepackage{amssymb}
\usepackage{slashed}
\usepackage{subfloat}
\usepackage{slashed}
\usepackage{color}
\usepackage{float}
\usepackage{multirow,multicol}
\usepackage{tabulary}
\usepackage[flushleft]{threeparttable}
\usepackage{pgfplots,subfigure}
\usepackage{xcolor}
\numberwithin{equation}{section}
\usepackage{tikz}
\usepackage{ulem}
\usepackage{hyperref}
\usepackage{nameref}
\usepackage{comment}

\usepgflibrary{arrows}
\usetikzlibrary{shapes.callouts}
\tikzset{
	level/.style   = { thick, },
	connect/.style = { dotted, red   },
	notice/.style  = { draw, rectangle callout, callout relative pointer={#1} },
	label/.style   = { text width=1cm }
}

\definecolor{acsblue}{RGB}{17,76,139}
\definecolor{shadecolor}{RGB}{255,241,204}
\usepackage{letltxmacro}
\makeatletter
\let\oldr@@t\r@@t
\def\r@@t#1#2{%
	\setbox0=\hbox{$\oldr@@t#1{#2\,}$}\dimen0=\ht0
	\advance\dimen0-0.2\ht0
	\setbox2=\hbox{\vrule height\ht0 depth -\dimen0}%
	{\box0\lower0.4pt\box2}}
\LetLtxMacro{\oldsqrt}{\sqrt}
\renewcommand*{\sqrt}[2][\ ]{\oldsqrt[#1]{#2}}
\makeatother
\usepackage{environ}

\begin{document}

\newcommand{{\ri}}{{\rm{i}}}
\newcommand{{\Psibar}}{{\bar{\Psi}}}
\newcommand*\var{\mathit}

\fontsize{8}{9}\selectfont

\title{\mdseries{On the Klein-Gordon bosonic fields in the Bonnor-Melvin spacetime with a cosmological constant in rainbow gravity: Bonnor-Melvin Domain Walls}}

\author{ \textit {\mdseries{Omar Mustafa}}$^{\ 1}$\footnote{\textit{ E-mail: omar.mustafa@emu.edu.tr (Corr. Author)} }~,~ \textit {\mdseries{Abdullah Guvendi}}$^{\ 2}$\footnote{\textit{E-mail: abdullah.guvendi@erzurum.edu.tr } }  \\
	\small \textit {$^{\ 1}$ \footnotesize Department of Physics, Eastern Mediterranean University, 99628, G. Magusa, north Cyprus, Mersin 10 - Türkiye}\\
	\small \textit {$^{\ 2}$\footnotesize  Department of Basic Sciences, Erzurum Technical University, 25050, Erzurum, Türkiye}}
\date{}
\maketitle

\begin{abstract}
We investigate the effect of rainbow gravity on Klein-Gordon (KG) bosons in the background of the magnetized Bonnor-Melvin (BM) spacetime with a cosmological constant. We first show that the very existence of the sinusoidal term \(\sin^2(\sqrt{2\Lambda}r)\), in the BM space-time metric, suggests that \(\sin^2(\sqrt{2\Lambda}r) \in [0,1],\) which consequently restricts the range of the radial coordinate \(r\) to \(r \in [0,\pi/\sqrt{2\Lambda}]\). Moreover, we show that at \(r = 0\) and \(r = \pi/\sqrt{2\Lambda}\), the magnetized BM-spacetime introduces domain walls (infinitely impenetrable hard walls) within which the KG bosonic fields are allowed to move. Interestingly, the magnetized BM-spacetime introduces not only two domain walls but a series of domain walls. However, we focus on the range \(r \in [0,\pi/\sqrt{2\Lambda}]\). A quantum particle remains indefinitely confined within this range and cannot be found elsewhere. Based on these findings, we report the effects of rainbow gravity on KG bosonic fields in BM-spacetime. We use three pairs of rainbow functions: \( f(\chi) = \frac{1}{1 - \tilde{\beta} |E|}, \, h(\chi) = 1 \); \( f(\chi) = (1 - \tilde{\beta} |E|)^{-1}, \, h(\chi) = 1 \); and \( f(\chi) = 1, \, h(\chi) = \sqrt{1 - \tilde{\beta} |E|^\upsilon} \), with \(\upsilon = 1,2\). Here, \(\chi = |E| / E_p\), \(\tilde{\beta} = \beta / E_p\), and \(\beta\) is the rainbow parameter. We found that while the pairs \((f,h)\) in the first and third cases fully comply with the theory of rainbow gravity and ensure that \(E_p\) is the maximum possible energy for particles and antiparticles, the second pair does not show any response to the effects of rainbow gravity. Moreover, the fascinating properties of this magnetized spacetime background can be useful for modeling magnetic domain walls in condensed matter systems. We show that the corresponding bosonic states can form magnetized, spinning vortices in monolayer materials, and these vortices can be driven by adjusting an out-of-plane aligned magnetic field.
\end{abstract}

\begin{small}
\begin{center}
\end{center}
\end{small}



\section{\mdseries{Introduction}}\label{sec:1}

From stars and accretion disks to galactic nuclei and intergalactic regions, magnetic fields have played a crucial role in a wide variety of phenomena of astrophysical interest. The study of these phenomena in the context of general relativity is inspired by their existence in the vicinity of compact massive objects in strong gravitational fields. The Bonner-Melvin (BM) universe, which describes a static cylindrically symmetric magnetic field (aligned with the symmetric axis) immersed in its own gravitational field, represents an interesting exact solution (among many admissible solutions) of the Einstein-Maxwell equation \cite{R1.1,R1.2,R1.3}. The magnetic field is known to contribute to the momentum-energy tensor, hence it curves the spacetime fabric and decreases away from the axis so that it does not collapse on itself. To counter this collapse and restore balance for a homogeneous field, a non-vanishing positive cosmological constant is incorporated into the solution. That is, the BM magnetic spacetime metric with a nonzero cosmological constant \(\Lambda > 0\),  expressed in units where \(\hbar = 1 = c\) \cite{R1.1}, reads  
\begin{equation}  
ds^{2} = -dt^{2} + dr^{2} + \alpha^{2} \sin^{2} \left( \sqrt{2\Lambda} \, r \right) d\varphi^{2} + dz^{2},  
\label{I.1}  
\end{equation}  
where  \(\Lambda\) has dimensions of inverse length squared, and the magnetic field is given by \(H = \sqrt{\Lambda} \, \alpha \sin \left( \sqrt{2\Lambda} \, r \right)\). Here, $\alpha $ is an integration constant \cite{R1.1} and represents a cosmic string parameter related to the deficit angle of the conical spacetime where $0<\alpha^2=1-\eta/2\pi<1$, with $\eta$ being the linear mass density of the cosmic string \cite{R1.4,R1.4.1}. \v{Z}ofka \cite{R1.1} has reported that when $\sqrt{2\Lambda }r=\pi $, the circumference of the rings $r=const.$ vanishes, suggesting that this is the location of an axis of some sort. In the current methodical proposal, we shall show that this is the exact location of an infinite hard wall (i.e., domain wall, one of the topological defects), as we investigate Klein-Gordon (KG) test particles/antiparticles in such a BM-spacetime with a cosmological constant. 

\vspace{0.15cm}
\setlength{\parindent}{0pt}

In fact, the grand unified theory has predicted some topological defects in the spacetime fabric \cite{R1.5,R1.6,R1.7,R1.8}, which include, but are not limited to, domain walls \cite{R1.6,R1.7}. Topological defects are known to modify the spectroscopic structure and dynamics of quantum mechanical particles \cite{R1.9,R1.10,R1.11,R1.12,R1.13,R1.14,R1.15,R1.16,R1.17,R1.18,AH,AY,AO,R1.19,R1.20,R1.21}. The exploration of the intricate and intriguing effects of gravitational fields on quantum mechanical systems forms a strong stimulus and motivation to investigate different quantum mechanical systems in the background of different spacetime fabrics. The BM-spacetime \cite{R1.1,R1.2,R1.3,R1.22,R1.23,AO-2,AAS} is one of such spacetime fabrics.

\vspace{0.15cm}
\setlength{\parindent}{0pt}

On the other hand, in rainbow gravity (RG), the energy of the probe particle is known to affect the spacetime background in the ultra-high-energy regime (i.e., ultraviolet regime, e.g., (cf., e.g., \cite{R1.24,R1.25,R1.26,R1.27,R1.28,R1.29,R1.30})) so that it modifies the standard relativistic energy-momentum dispersion relation (MDR) in the ultraviolet regime into%
\begin{equation}
E^{2}f\left( \chi\right) ^{2}-p^{2}h(\chi)
^{2}=m_\circ^{2}, \label{I.2}
\end{equation}%
which is in fact a common suggestion of most approaches to quantum gravity (such as string field theory, loop quantum gravity, and non-commutative geometry \cite{R1.31,R1.32,R1.33}, respectively). Consequently, the BM-spacetime background under rainbow gravity is modified to read
\begin{equation}
ds^{2}=-\frac{dt^{2}}{f\left( \chi\right) ^{2}}+\frac{1}{h\left( \chi\right) ^{2}}\left(dr^{2}+\alpha ^{2}\sin ^{2}\left( \sqrt{2\Lambda }r\right)
\,d\varphi ^{2}+dz^{2}\right),  \label{I.3}
\end{equation}%
where $f(\chi)$ and $h(\chi)$ are called rainbow functions that satisfy $\lim\limits_{\chi\rightarrow 0}f\left( \chi\right) =1=\lim\limits_{\chi\rightarrow 0}h\left( \chi\right)$, with $0\leq\chi=|E|/E_{p}\leq 1$ (to retrieve the standard energy-momentum dispersion relation in the infrared regime). At this point, one should notice that $\chi=|E|/E_p$ is a fine-tuned rainbow gravity parameter that allows rainbow gravity to affect relativistic particles and antiparticles alike (e.g., \cite{R1.30,R1.34,R1.35}). Moreover, the rainbow functions are chosen so that they secure the Planck energy $E_p$ as a threshold between quantum and classical descriptions and establish another invariant in addition to the speed of light. 

\vspace{0.15cm}
\setlength{\parindent}{0pt}

Very recently, the effects of rainbow gravity on scalar bosonic and oscillator fields in BM-spacetime, as described in (\ref{I.3}), have been investigated \cite{R1.4}. The study employed the Magueijo-Smolin rainbow function pair \( f\left( \chi \right) =1/\left( 1-\tilde{\beta}\left\vert E\right\vert \right) \), \( h\left( \chi \right) =1 \), developed in the context of the varying speed of light hypothesis \cite{R1.36}, as well as an alternative pair, \( f\left( \chi \right) = (1-\tilde{\beta}\left\vert E\right\vert )^{-1} = h\left( \chi \right) \), introduced in \cite{R1.24}. While the latter does not lead to a varying speed of light, it has been considered a possible resolution to the horizon problem. Here, \( \tilde{\beta}=\beta /E_{p} \), where \( \beta \) is the rainbow parameter. In this work, we undertake a thorough reexamination of the problem and meticulously discuss and report our points of view on this issue. Our objective is to derive exact solutions with well-defined boundary conditions for Klein-Gordon quantum particles and antiparticles in the gravitational field of a magnetized BM-spacetime background within the framework of rainbow gravity (\ref{I.3}).

\vspace{0.15cm}
\setlength{\parindent}{0pt}

The very existence of the sinusoidal term, \(\sin(\sqrt{2\Lambda}r)^2\), in the BM-spacetime metric (\ref{I.3}) suggests that \(\sin(\sqrt{2\Lambda}r)^2\in[0,1]\) and consequently introduces some restrictions on the allowed range of the radial coordinate \(r\), so \(\sqrt{2\Lambda}r=\kappa\pi;\,\kappa\in\mathbb{Z} \Rightarrow r=\kappa\pi/\sqrt{2\Lambda};\,\kappa=0,1,2,\cdots\).  Having the radial coordinate origin at \(r=0\), for \(\kappa=0\), would necessarily suggest that \(\kappa=1\Rightarrow r=\pi/\sqrt{2\Lambda}\) is the location of the upper limit for the range of $r$ so that \(r\in[0,\pi/\sqrt{2\Lambda}]\). The upper bound indicates that the system's dynamics are profoundly determined by the cosmological constant. This would in fact agree with \v{Z}ofka's \cite{R1.1} proposal that $\sqrt{2\Lambda }r=\pi$  is the location of an axis of some sort. In Section \ref{sec:2}, we shall show that $\sqrt{2\Lambda }r=\pi$  is the location of a domain wall (infinite impenetrable wall) that restricts the motion of particles within the range \(r\in[0,\pi/\sqrt{2\Lambda}]\).  We show this through some analysis of the effective potential introduced by the gravitational field of the BM-spacetime. That is, we shall not follow the commonly used assumption that \(r<<1\rightarrow \sin(r)\sim r\Rightarrow \tan(r)\sim r\) (e.g. \cite{R1.4,R1.4.1}). Such an assumption, in our opinion, is not valid because it not only changes the dynamics of the quantum mechanical system at hand, but also eliminates the domain wall introduced by the BM-spacetime. The effect of rainbow gravity is discussed in Section \ref{sec:3}, where we use three pairs of rainbow functions. In addition to the two rainbow functions mentioned above, \(f\left( \chi \right) =1/\left( 1-\tilde{\beta}\left\vert E\right\vert \right),\,h(\chi)=1\) and  $f\left( \chi \right) =$ $\left(
1-\tilde{\beta}\left\vert E\right\vert \right) ^{-1}=$ $h\left( \chi \right)$, we also use the loop quantum gravity-motivated pairs \cite{R1.38,R1.39},  $f(\chi)=1$, $h\left(\chi \right) = \sqrt{1-%
\tilde{\beta}(\left\vert E\right\vert)^\upsilon}$\,; $\upsilon=1,2$.  The latter pairs are found to completely comply with the rainbow gravity theory because they ensure that the Planck energy $E_p$ is the maximum possible energy for particles and antiparticles (e.g., \cite{R1.34,R1.35,R1.40}). We conclude in Section \ref{sec:4}.

\section{\mdseries{Preliminary analysis of the effective gravitational potential}}\label{sec:2}

\begin{figure*}[ht!]
\centering
\includegraphics[width=0.25\textwidth]{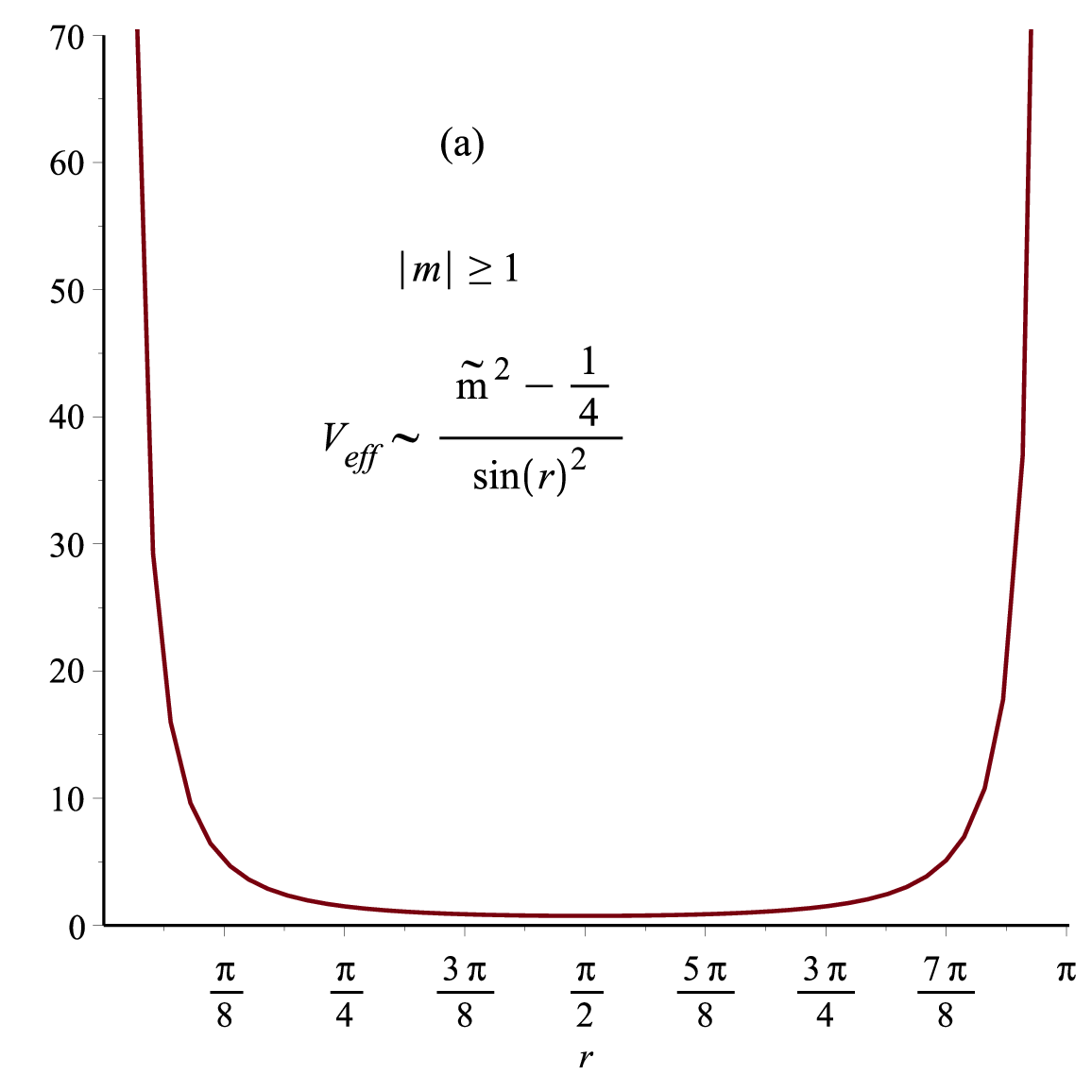}
\includegraphics[width=0.25\textwidth]{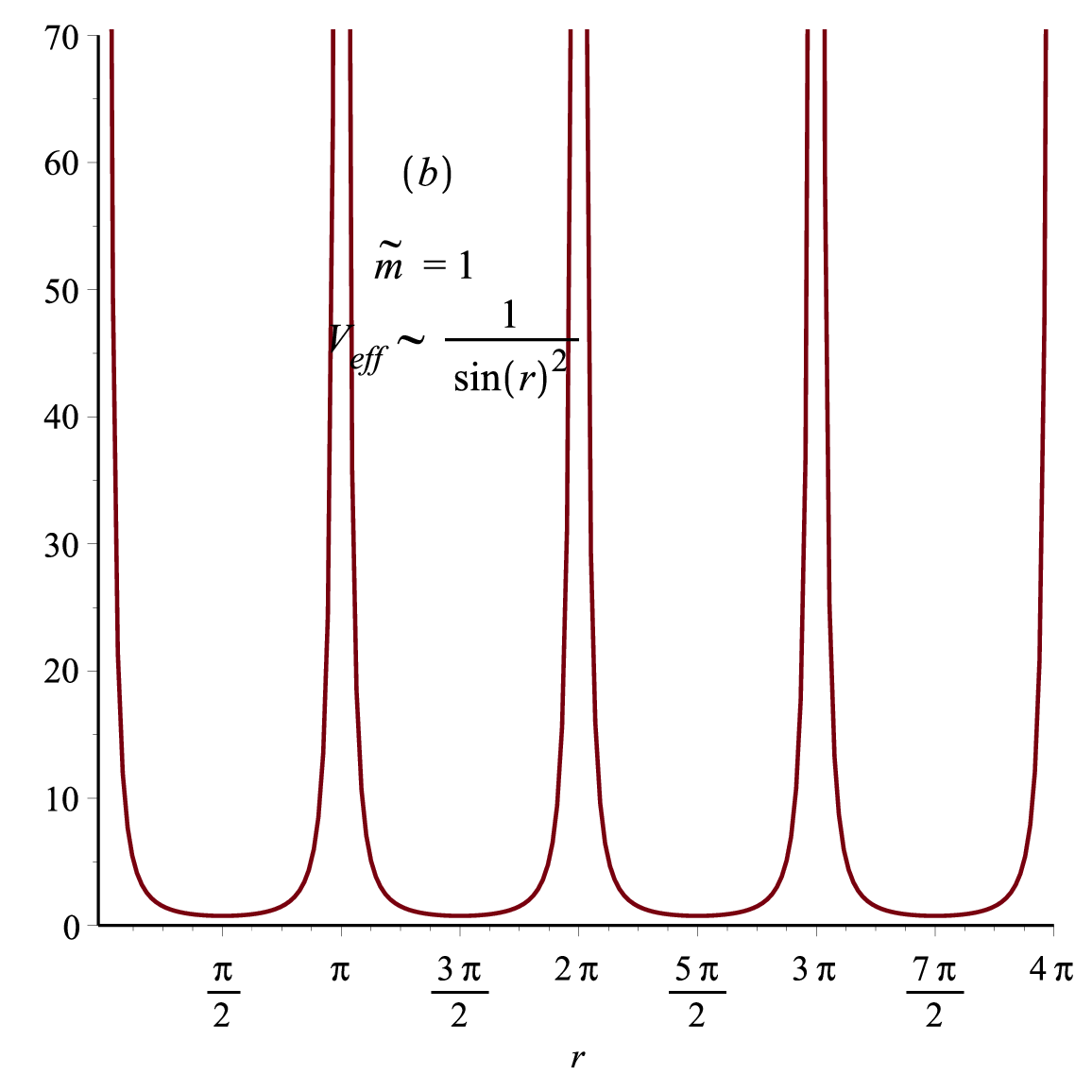}
\includegraphics[width=0.25\textwidth]{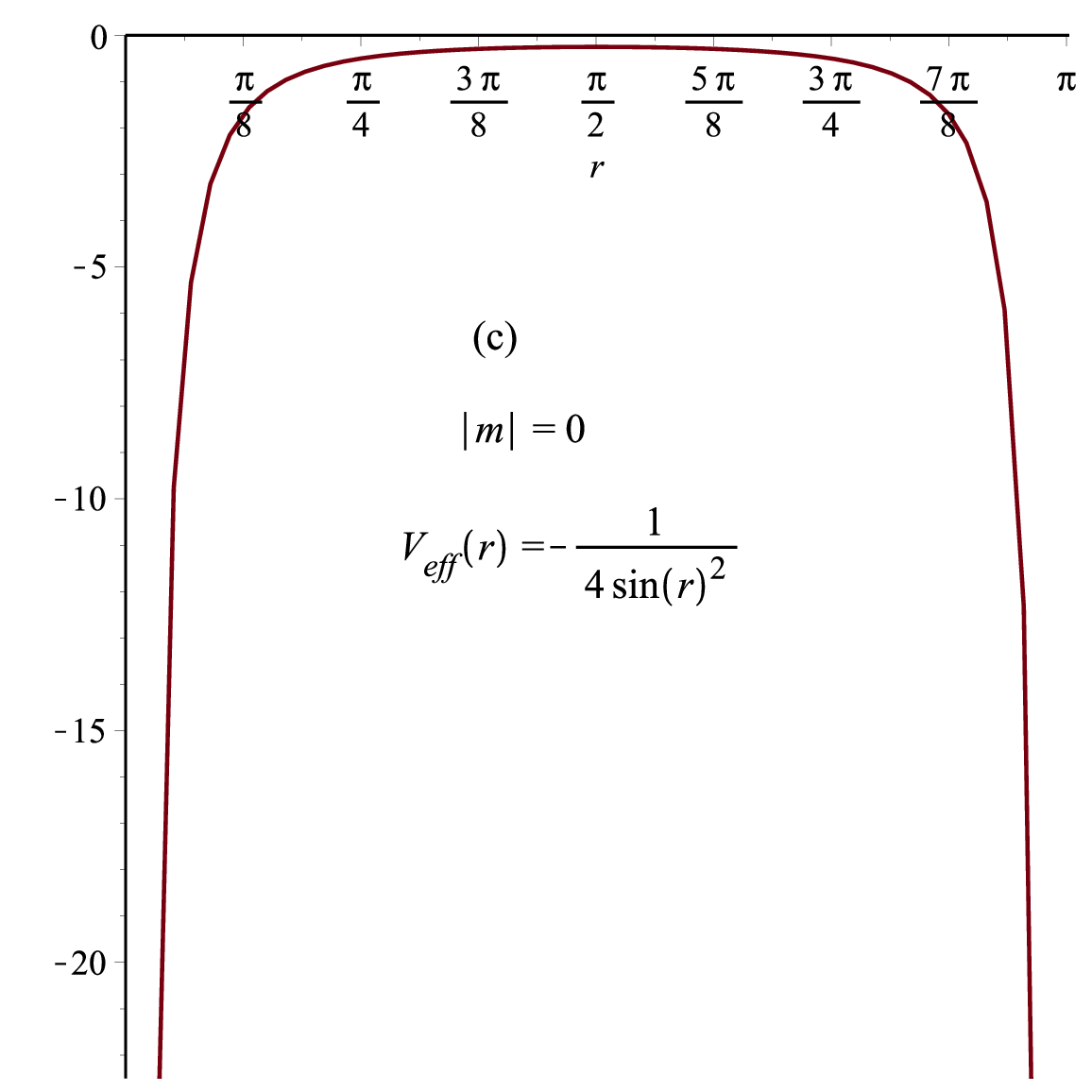}
\includegraphics[width=0.25\textwidth]{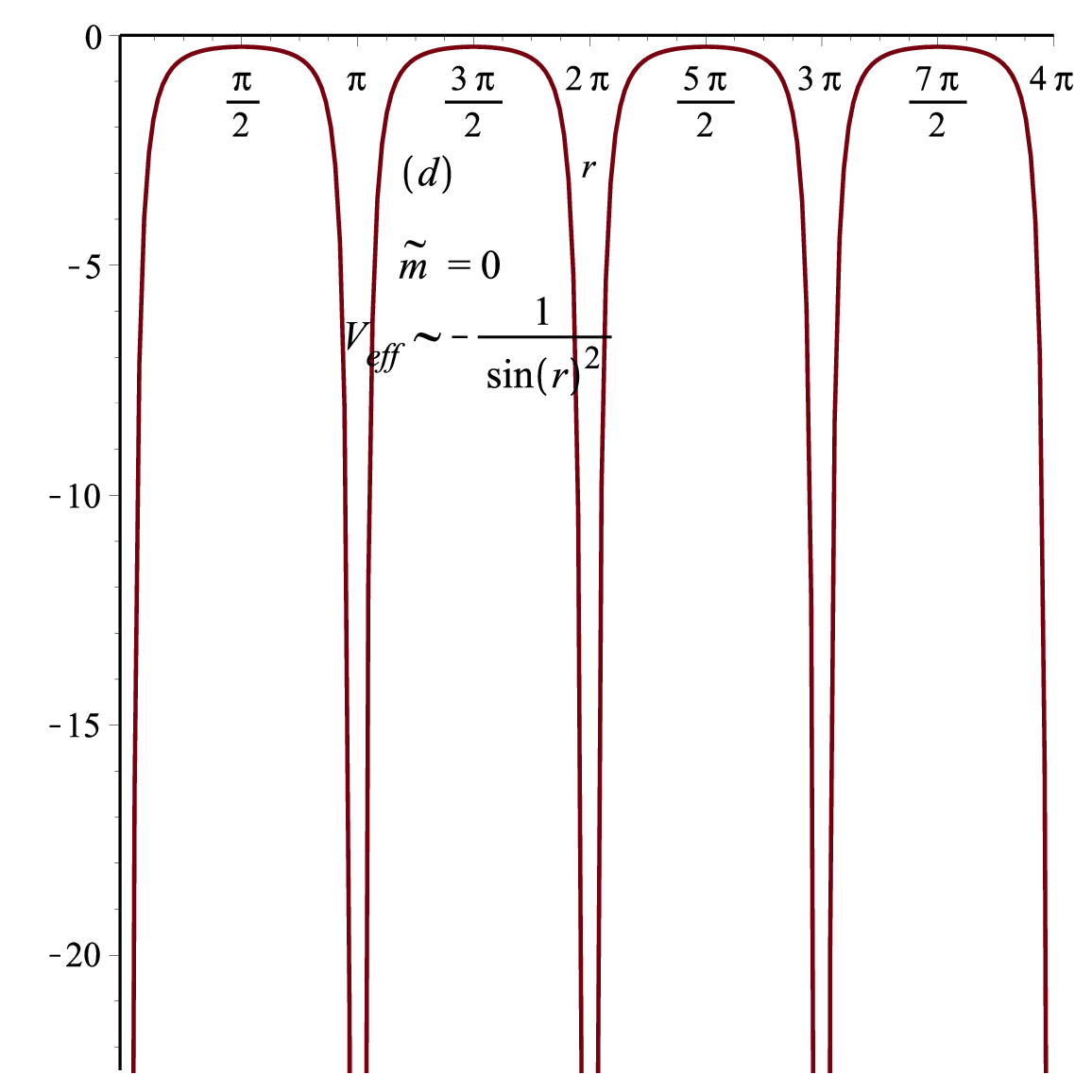}
\includegraphics[width=0.25\textwidth]{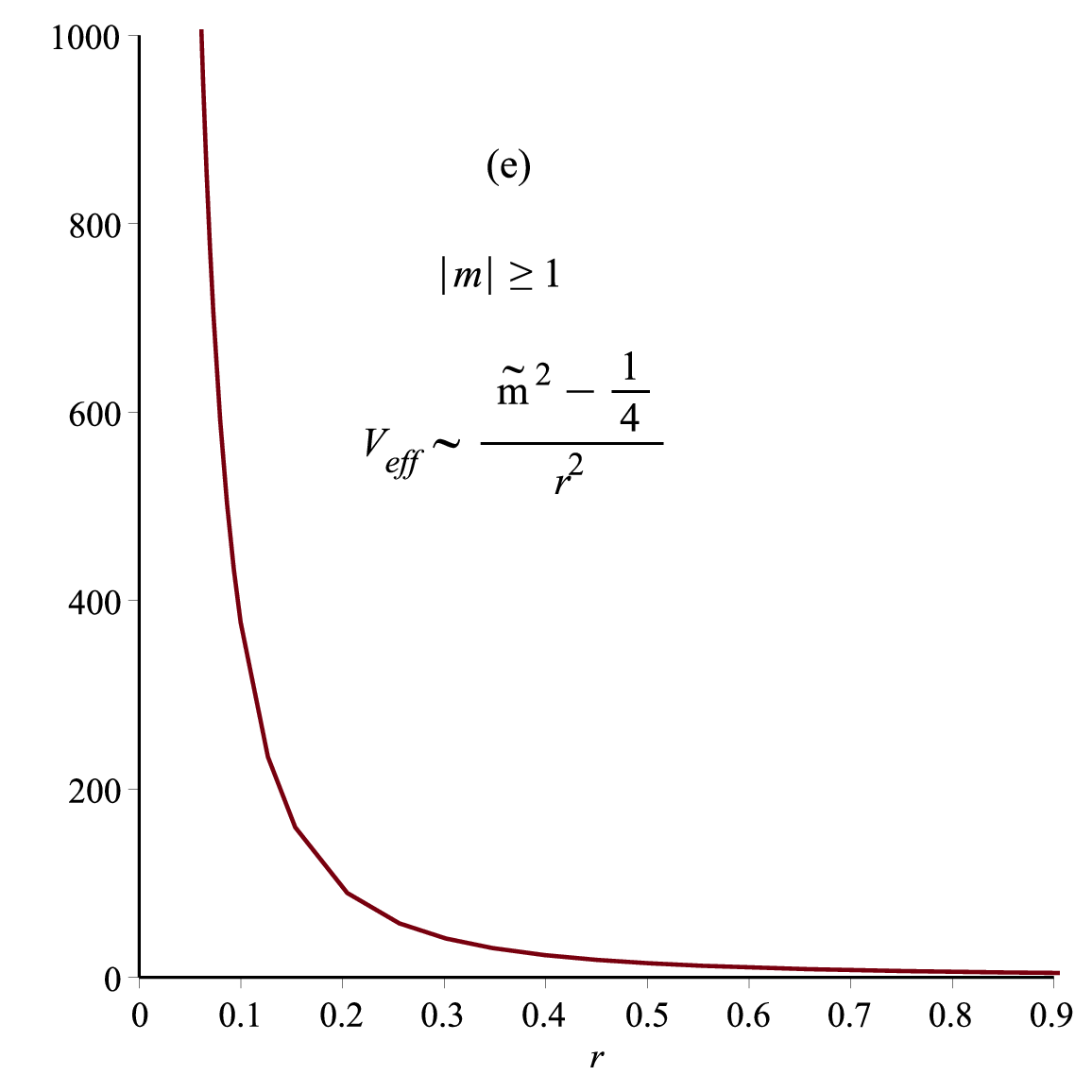}
\includegraphics[width=0.25\textwidth]{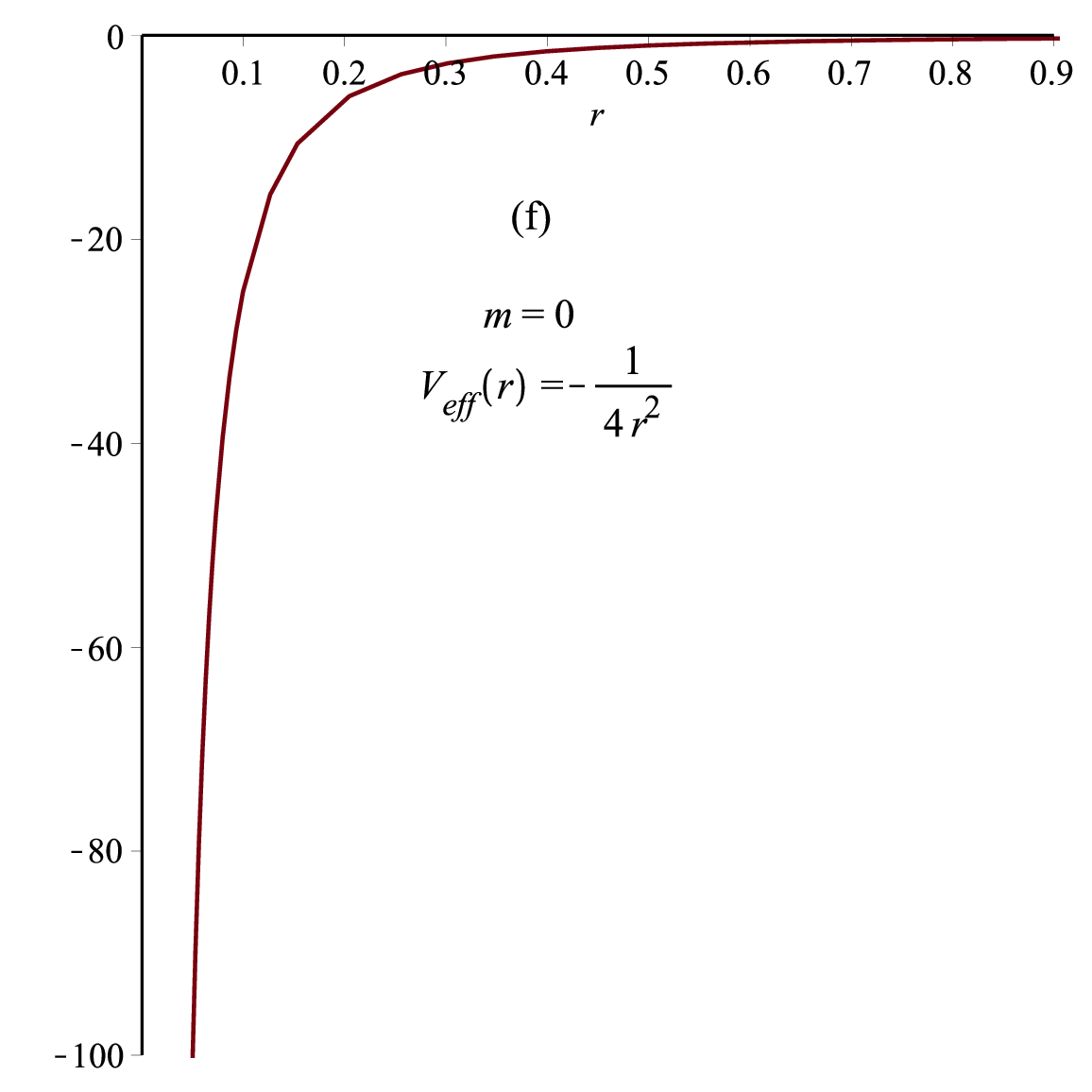}
\caption{\footnotesize  (a) Effective potential (\ref{II.4}) for the magnetic quantum number $|m|\geq 1$ with $r\in[0,\pi]$, (b) for $\tilde{m}=1$ with $r\in[0,4\pi]$, (c) for $m=0$ with $r\in[0,\pi]$, (d) for $m=0$ with $r\in[0,4\pi]$, whereas the effective potential (\ref{II.4}) with the asymptotic limit $r<<1\Rightarrow \sin(r)=r=\tan(r)$ (used in \cite{R1.4}) (e) for $|m|\geq 1$, and (f) for $m=0$.}
\label{fig1}
\end{figure*}

We start (as in \cite{R1.4}) with rescaling the coordinates in (\ref{I.3}) so that $\sqrt{2\Lambda}r\rightarrow r$ and $\sqrt{2\Lambda}\varphi\rightarrow \varphi$ so that we can rewrite the BM-spacetime in the RG metric as
\begin{equation}
ds^{2}=-\frac{dt^{2}}{f\left( \chi\right) ^{2}}+\frac{1}{h\left( \chi\right) ^{2}}\left(dz^{2}+\frac{1}{2\Lambda}\left[dr^{2}+\alpha ^{2}\sin ^{2}\left( r\right)
\,d\varphi ^{2}\right]\right).  \label{I.4}
\end{equation}%
For which the contravariant metric tensor $g^{\mu\nu}$ has the non-vanishing elements
\begin{equation}
    g^{00}=-f(\chi)^2,\,\,g^{11}=2\Lambda h(\chi)^2,\,\,g^{22}=\frac{2\Lambda h(\chi)^2}{\alpha^2\sin(r)^2},\,\,g^{33}=h(\chi)^2, \label{I.5}
\end{equation}%
and the determinant of the covariant metric tensor is given by $$det(g_{\mu\nu})=g=-\frac{\alpha^2 \sin(r)^2}{ 4\Lambda^2 h(\chi)^6 f(\chi)^2}.$$
Then, the KG-equation
\begin{equation}
    \frac{1}{\sqrt{-g}}\partial_\mu \,\sqrt{-g} \, g^{\mu\nu}\,\partial_\nu \, \psi(t,r,\varphi,z)=M^2\, \psi(t,r,\varphi,z),\label{I.6}
\end{equation}%
where $M=m_\circ$ is the rest mass energy (also called the rest energy). We now use the substitution $$\psi(t,r,\varphi,z)=e^{i(m\varphi+k\,z-Et)} \psi(r)$$ to eventually yield
\begin{equation}
\psi ^{\prime \prime }\left( r\right) +\frac{1}{\tan \left( r\right) }\psi
^{\prime }\left( r\right) +\left( \lambda -\frac{\tilde{m}^{2}}{\sin \left(
r\right) ^{2}}\right) \psi \left( r\right) =0,  \label{II.1}
\end{equation}
with
\begin{equation}
\tilde{m}=\frac{m}{\alpha },\text{ and }\;\lambda =\frac{f\left( \chi
\right) ^{2}E^{2}-M^{2}}{ 2\Lambda h\left( \chi \right) ^{2}}-\frac{k^{2}}{2\Lambda},  \label{II.2}
\end{equation}
where $\tilde{m}=\iota$ used in \cite{R1.4} is the magnetic quantum number. For more details on (\ref{II.1}), the reader is advised to visit Sections 1 and 2 of \cite{R1.4}.

At this point, we wish to elaborate on the effective potential governing the motion of KG particles in 
BM-spacetime and discuss the validity of the range of radial coordinate \(r\). To do so, we first need to remove the first-order derivative using the substitution 
$$\psi \left( r\right) =\frac{U\left( r\right)}{\sqrt{\sin \left( r\right)}}.$$Which in turn should satisfy the regular textbook conditions that \(\psi(r)\) and \(U(r)\) are finite everywhere and at least \(U\left( 0\right) =0\). This would allow us to obtain
\begin{equation}
U^{\prime \prime }\left( r\right) +\left( \tilde{\lambda}-\frac{\left( 
\tilde{m}^{2}-1/4\right) }{\sin \left(r\right)^{2}}\right) U\left(
r\right) =0,\;\tilde{\lambda}=\lambda +\frac{1}{4}.  \label{II.3}
\end{equation}
This equation resembles the one-dimensional form of the radial Schr\"{o}dinger equation with an effective radial potential given by
\begin{equation}
V_{eff}\left( r\right) =\frac{\left( \tilde{m}^{2}-1/4\right) }{\sin \left(
r\right) ^{2}}.  \label{II.4}
\end{equation}
This effective potential exhibits distinct behaviors depending on the value of \( r \) and the magnetic quantum number \( m \), as documented in Fig. \ref{fig1}(a) and Fig. \ref{fig1}(b). It is obvious that this effective potential function is characterized by singularities at \( r = \kappa\pi;\, \mathbb{Z}\ni\kappa=0,\pi,2\pi,\cdots \) (also clearly observed in Fig. \ref{fig1}(a) and Fig. \ref{fig1}(b)). One should also observe that while this effective potential manifestly introduces a repulsive core for \(\left( 
\tilde{m}^{2}-1/4\right) >0\Longrightarrow \left\vert m\right\vert \geq 1\), (with a repulsive gravitational force field that increases as \( |m| \) increases), it becomes attractive for \(m=0\) (and extremely attractive for \(r<<1\), as is obvious in Figs. \ref{fig1}(c) and \ref{fig1}(d)). Moreover, since \(\tilde{\lambda}\equiv\tilde{\lambda}(E^2)\) is the corresponding eigenvalue for the one-dimensional Schr\"{o}dinger-like KG equation (\ref{II.3}), one should require that \(\tilde{\lambda}\equiv\tilde{\lambda}(E^2)>0\) in order to be able to account for KG particles, \(E=E_+=+|E|\), and antiparticles \(E=E_-=-|E|\), bound states. However, for the eigenvalues \(\tilde{\lambda}\equiv\tilde{\lambda}(E^2)<0\) (i.e., the only allowed eigenvalues for \(m=0\), as is clear from Fig. \ref{fig1}(c) and \ref{fig1}(d)), we have
\begin{equation}
   \tilde{\lambda}\equiv \tilde{\lambda}\left( E^{2}\right)<0\Rightarrow \tilde{\lambda}= \frac{E^2}{2\Lambda}+\frac{1}{4}\Rightarrow E^2=-2\Lambda \left(|\tilde{\lambda}|+\frac{1}{4}\right), \label{II.4.1}
\end{equation}
using (\ref{II.2}) and (\ref{II.3}) for massless KG-particle/antiparticle in no rainbow gravity, $f(\chi)=1=h(\chi)$, and $k=0$. Therefore, we have \(E=\pm i\omega=\pm i\sqrt{2\Lambda \left(|\tilde{\lambda}|+\frac{1}{4}\right)}\), and the corresponding states cannot be steady states because these modes grow or decay over time since \(\psi \propto e^{-iE t}\) (with decay time \(\tau=\frac{1}{|\Im E|}\)). In this study, we are only interested in KG particles/antiparticles that are confined to move within the impenetrable infinite domain walls (i.e., \(r\in[0,\pi]\)) generated by the BM-spacetime, and carrying the magnetic quantum numbers \(m=\pm1,\pm2,\cdots\). Furthermore, the effective potential (\ref{II.4}) suggests that, for \(m\geq1\), multiple domain walls at \(r = \kappa\pi\) (within which KG particles and antiparticles are destined/allowed to move) are manifestly introduced by the gravitational field produced by the magnetized BM-spacetime. Yet, the quantum mechanical solution within the range \(r\in[0,\pi]\) would be the same solution for \(r\in[\pi,2\pi],\,r\in[2\pi,3\pi],\, \cdots\), as we shall witness in the following section. However, a quantum particle moving within \(r\in[0,\pi]\) is indefinitely trapped therein as a consequence of the domain walls introduced by the BM-spacetime fabric. Therefore, it is unlikely for this quantum particle to be found moving within the subsequent domain walls governing the radial ranges \(r\in[\pi,2\pi],\, r\in[2\pi,3\pi], \cdots\). In the current study, therefore, we focus on the KG-particles in BM-spacetime moving within \(r\in[0,\pi]\).

On the other hand, it is unavoidably necessary and vital to discuss the commonly used approximation assumption (e.g., \cite{R1.4,R1.4.1}) that \(r\ll1\Longrightarrow \sin (r)\sim r\) and \(\tan \left( r\right) \sim r\). This assumption would result in an effective potential \(V_{eff}\left( r\right) =\left( \tilde{m}%
^{2}-1/4\right) /r^{2}\) as shown in Figs. \ref{fig1}(e) and \ref{fig1}(f). It is clear that none of the effective potentials in 1(e) and 1(f) can support KG bound states because of the arguments we have discussed above. This kind of approximation would effectively transform the effective potential (\ref{II.4}) that supports bound states (without those for \(m=0\)) into an effective potential that does not support any bound states (in the one-dimensional quantum mechanical textbook language, so to speak). One should, therefore, feel very reluctant to suggest a hard wall at some \(r=r_{\circ},\) e.g., \cite{R1.4,R1.4.1}, (a hard wall that is not manifestly introduced by the gravitational field of the magnetized BM-spacetime) and used the asymptotic form of Bessel functions of the first kind,
\begin{equation}
J_{\iota }\left( r\,\Theta \right) \underset{ r\,\Theta\rightarrow \infty }{\sim }%
\cos \left( r\,\Theta -\frac{\iota \pi }{2}-\frac{\pi }{4}\right) ,
\label{II.5}
\end{equation}%
where their \(\Theta =\sqrt{\lambda }\) and \(\iota =\tilde{m}\) above, as clearly stated by Abramowitz and Stegun \cite{R1.37}. Here, we should observe that this asymptotic form is valid for \(r\rightarrow \infty\). This would not only immediately contradict the initial assumption that \(r\ll1\Longrightarrow \sin (r)\sim r\), but also change the physically allowed range of validity of the radial coordinate form \(0\leq r\leq \pi\) to \(0\leq r\ll1\). Moreover, the assumption of a hard wall (i.e., domain wall) at some \(r=r_{\circ}<<1\) is not manifestly introduced by the BM-spacetime fabric but rather a hypothetical unrealistic one. It is therefore interesting to meticulously investigate rainbow gravity effects on KG particles/antiparticles in BM-spacetime with cosmological constant. We do so in the sequel.\\

\section{\mdseries{Rainbow gravity effects}}\label{sec:3}

\begin{figure*}[ht!]
\centering
\includegraphics[width=0.25\textwidth]{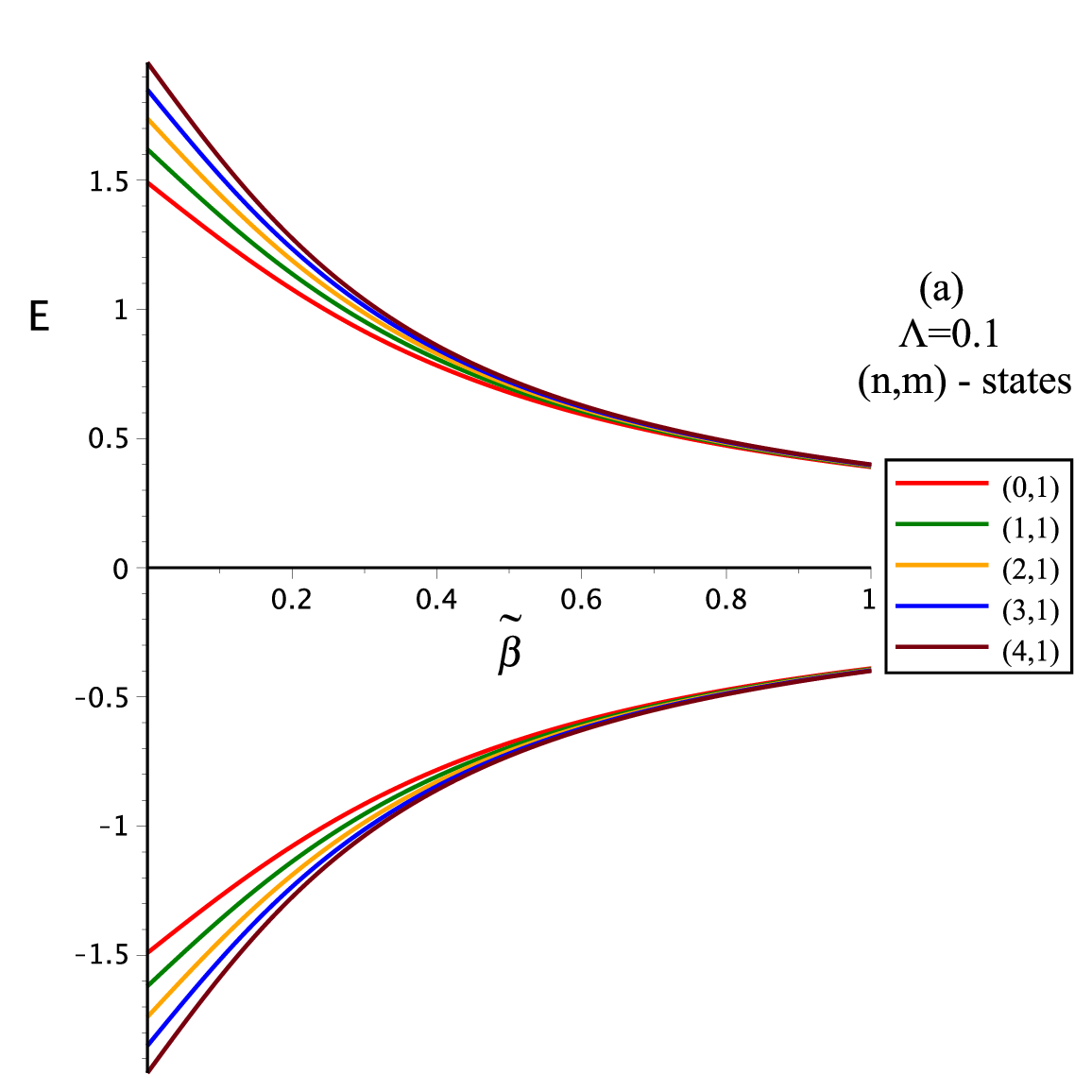}
\includegraphics[width=0.25\textwidth]{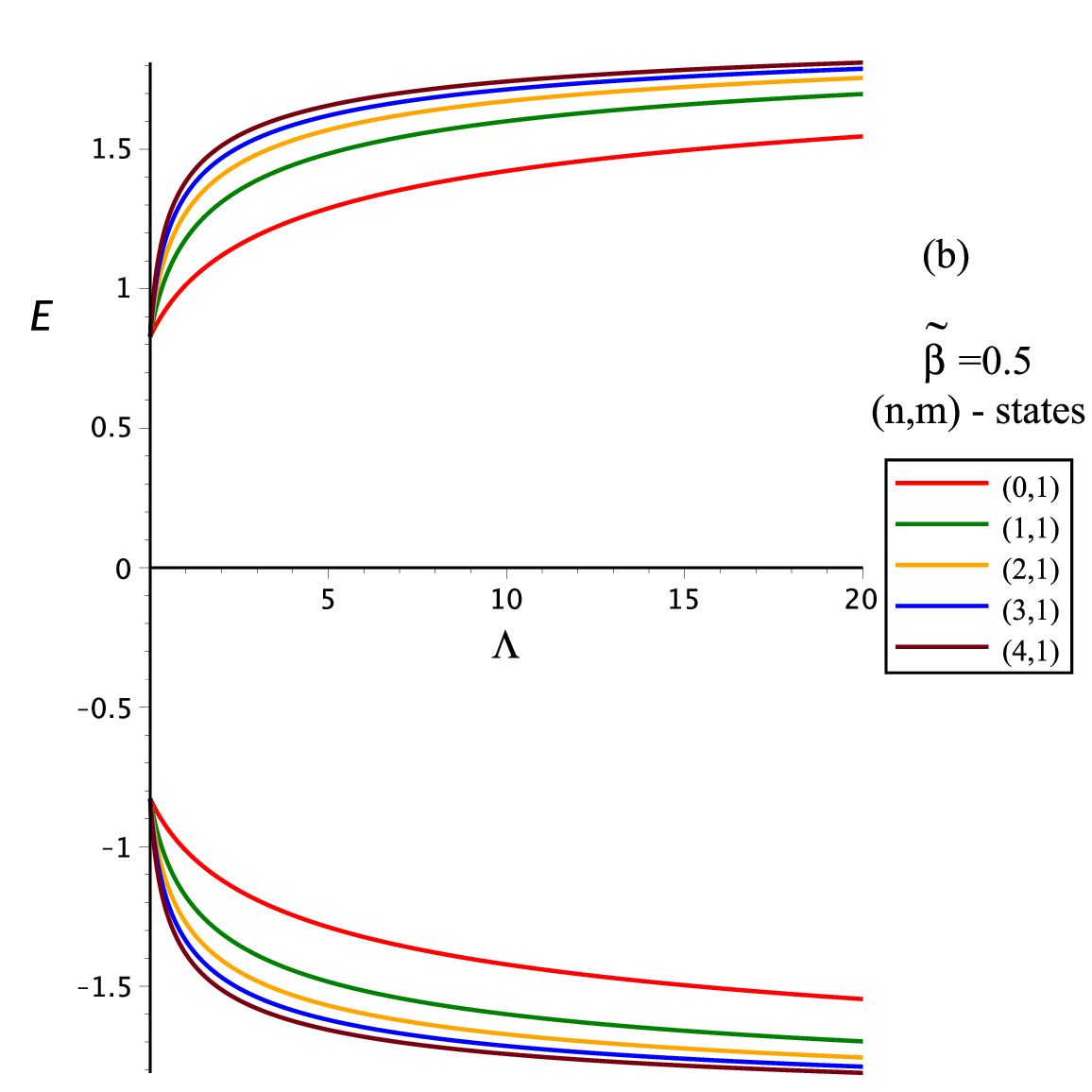}
\includegraphics[width=0.25\textwidth]{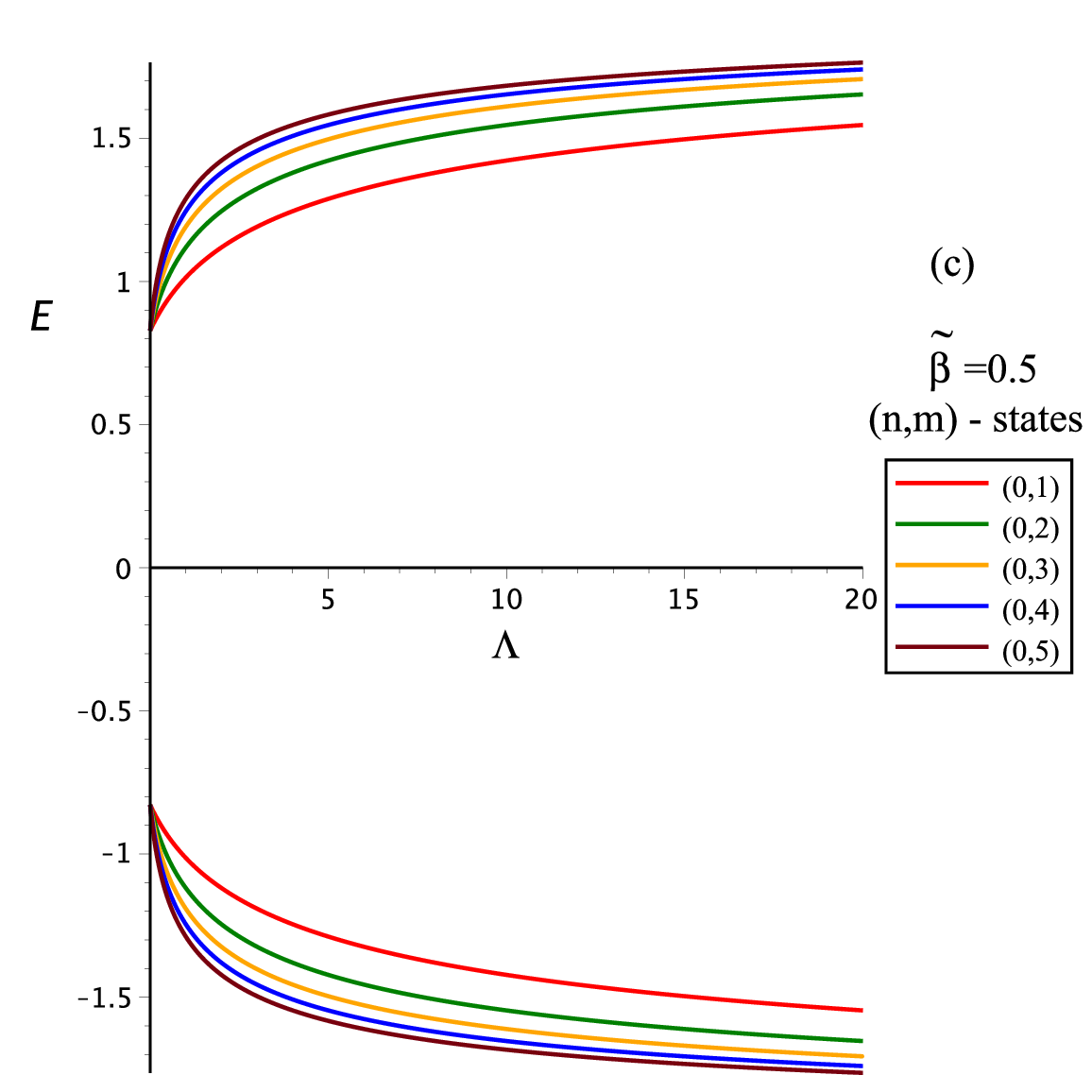}
\caption{\footnotesize  The figure shows the energy levels for KG particles and antiparticles given by (\ref{III.1.6}) so that we plot (a) $E$ against $\tilde{\beta}$ for $m=1$ and $n=0,1,2,3,4$, (b) $E$ against the cosmological constant $\Lambda$ for $m=1$, $n=0,1,2,3,4$ and $\tilde{\beta}=0.5$, and (c) $E$ against the cosmological constant $\Lambda$ for $n=0$, $m=1,2,3,4,5$ and $\tilde{\beta}=0.5$.}
\label{fig2}
\end{figure*}

In this section, we shall use the effective potential in (\ref{II.4}) and elaborate on the validity of the solution for the one-dimensional Schrödinger-like radial equation (\ref{II.3}) above. In so doing, we use the first change of variable \(x = \sin(r)\) to obtain
\begin{equation}
\left( x^{2}-1\right) U^{\prime \prime }\left( x\right) +xU^{\prime }\left(
r\right) +\left[ \frac{\tilde{m}^{2}-1/4}{x^{2}}-\tilde{\lambda}\right]
U\left( x\right) =0.  \label{III.6}
\end{equation}
Using yet another change of variables \(y = x^{2}\) we obtain
\begin{equation}
\left( y^{2}-y\right) U^{\prime \prime }\left( y\right) +\left( y-\frac{1}{2}%
\right) U^{\prime }\left( r\right) +\left[ \frac{\tilde{M}}{y}-\frac{\tilde{%
\lambda}}{4}\right] U\left( y\right) =0,  \label{III.7}
\end{equation}
where 
\begin{equation}
\tilde{M}=\frac{\tilde{m}^{2}}{4}-\frac{1}{16}.  \label{III.71}
\end{equation}
The power series solution
\begin{equation}
U\left( y\right) =\sum\limits_{j=0}^{\infty }A_{j}\,y^{j+\sigma }
\label{III.8}
\end{equation}
of which, when substituted into (\ref{III.7}), would eventually yield
\begin{equation}
\begin{split}
&\sum\limits_{j=0}^{\infty }A_{j}\left[ \left( j+\sigma \right) ^{2}-\frac{%
\tilde{\lambda}}{4}\right] \,y^{j+\sigma }\\
&+\sum\limits_{j=0}^{\infty }A_{j+1}\left[ \tilde{M}-\left( j+\sigma +1\right) \left( j+\sigma +\frac{1}{2}
\right) \right] \,y^{j+\sigma } \\
&+A_{0}\left[ \tilde{M}-\sigma \left( \sigma -\frac{1}{2}\right) \right]
\,y^{\sigma -1}=0.  \label{III.9}
\end{split}
\end{equation}
Next, \(A_{0}\neq 0\) would necessarily imply that
\begin{equation}
\tilde{M}-\sigma \left( \sigma -\frac{1}{2}\right) =0\Longrightarrow \sigma =%
\frac{1}{4}\pm \frac{\left\vert \tilde{m}\right\vert }{2}.  \label{III.10}
\end{equation}
Obviously, we shall adopt \(\sigma =\frac{1}{4}+\frac{\left\vert \tilde{m}%
\right\vert }{2}\) so that the radial function \(U\left( y\right)
\longrightarrow 0\) as \(y\longrightarrow 0\) for both limits \(r\in \left[
0,\pi \right] \Longrightarrow \sin \left( 0\right) =0=\sin \left( \pi
\right) \) (note that \(r\in \left[ 0,\pi \right] \) is the allowed range for the particle's motion in an infinite textbook-like potential well of width \(L=\pi \)). Hence, we obtain
\begin{gather}
A_{j}\left[ \left( j+\sigma \right) ^{2}-\frac{%
\tilde{\lambda}}{4}\right] +A_{j+1}%
\left[ \tilde{M}-\left( j+\sigma +1\right) \left( j+\sigma +\frac{1}{2}%
\right) \right]  =0,  \label{III.91}
\end{gather}
which would provide for \(j\geq 0\) the correlation between the coefficients of the power series in (\ref{III.8}). Moreover, we now truncate the power series to a polynomial of order \(n\geq 0\) by the requirement that \(\forall j=n\) we have \(A_{n+1}=0\) and \(A_{n}\neq 0\). The latter would imply that
\begin{equation}
\left( n+\sigma \right) ^{2}-\frac{\tilde{\lambda}}{4}=0\Longrightarrow 
\tilde{\lambda}=\left( 2n+\left\vert \tilde{m}\right\vert +\frac{1}{2}%
\right) ^{2}\,;\quad m=\pm1,\pm2,\cdots.  \label{III.11}
\end{equation}
Moreover, our \(U(y)\) now reads
\begin{equation}
\begin{split}
&  U\left( y\right) =\mathcal{C\,}
y^{\left\vert \tilde{m}\right\vert
/2+1/4}\sum\limits_{j=0}^{n}A_{j}\,y^{j}\\
&\Longrightarrow U\left( r\right) =
\mathcal{C}\,\sin \left( r\right) ^{\left\vert \tilde{m}\right\vert+1/2
}\sum\limits_{j=0}^{n}A_{j}\,y^{j},\label{III.11.1}
\end{split}
\end{equation}
and our radial wave function \(\psi \left( r\right) \) would, consequently, read
\begin{equation}
\begin{split}
&\psi \left( y\right) =\frac{U\left( y\right) }{y^{1/4}}=\mathcal{C\,}
y^{\left\vert \tilde{m}\right\vert
/2}\sum\limits_{j=0}^{n}A_{j}\,y^{j}\\
&\Longrightarrow \psi_{n,m} \left( r\right) =
\mathcal{C}\,\sin \left( r\right) ^{\left\vert \tilde{m}\right\vert
}\sum\limits_{j=0}^{n}A_{j}\,y^{j},  \label{III.12}
\end{split}
\end{equation}
where \(y=x^{2}=\sin \left( r\right) ^{2}\). At this point, it is interesting to mention that the polynomial in (\ref{III.11.1}) is a hypergeometric polynomial. However, it is obvious that our \(U(r)\), in (\ref{III.11.1}), satisfies the textbook conditions \(U(0)=0=U(\pi)\). Similarly, \(\psi(0)=0=\psi(\pi)\). The result (\ref{III.11}), along with (\ref{II.2}), (\ref{II.3}), and (\ref%
{III.71}), would imply that
\begin{equation}
\begin{split}
&f\left( \chi \right) ^{2}E^{2}=h\left( \chi \right) ^{2}\mathcal{G}
_{nm}+M^{2};\\
&\mathcal{G}_{nm}=2\Lambda \left( 2n+\left\vert \tilde{m}
\right\vert \right) \left( 2n+\left\vert \tilde{m}\right\vert +1\right)
+k^{2}.  \label{III.13}
\end{split}
\end{equation}
Notably,  in no rainbow gravity \(\beta=0\), one obtains
\begin{equation}
\begin{split}
&E_\pm=\pm\sqrt{\mathcal{\tilde{G}}_{nm}};\\
&\mathcal{\tilde{G}}_{nm}=\mathcal{G}_{nm}+M^2. \label{III.13.1}
\end{split}
\end{equation}
This result demonstrates that the magnetic domain walls, influenced by the cosmological constant \(\Lambda\), induce cyclotron motion in the particle, giving rise to quantum oscillatory behavior (as detailed in \cite{AO-2}), while preserving the symmetry of the energy spectrum around the zero energy, (\(E = 0\)). To investigate the effects of rainbow gravity on KG-test particles/antiparticles, three different pairs of rainbow functions are to be used here: (i) \(f\left( \chi \right) =\left( 1-\beta \chi \right) ^{-1}\), \(
h\left( \chi \right) =1\), (ii) \(f\left( \chi \right) = \left( 1-\beta
\chi \right) ^{-1} = h\left( \chi \right)\), and (iii) \(f(\chi)=1\), \(h\left(\chi \right) = \sqrt{1-
\beta\chi^{\upsilon}}\); \(\upsilon=1,2\), where \(0\leq \chi =E/E_{p}\leq
1\). At this point, one should observe the limitations imposed upon the allowed values of \(\chi\) that necessarily and mandatorily enforce the condition that \(0\leq \chi =\left\vert E\right\vert /E_{p}\leq 1\) (and not \(\chi =E/E_{p}\) to account for particles' and antiparticles' energies alike). This would in turn allow us to rewrite the rainbow functions pairs as (i) \(
f\left( \chi \right) =\left( 1-\tilde{\beta}\left\vert E\right\vert \right)
^{-1}\), \(h\left( \chi \right) =1\), (ii) \(f\left( \chi \right) = \left(
1-\tilde{\beta}\left\vert E\right\vert \right) ^{-1} = h\left( \chi \right)\), and (iii) \(f(\chi)=1\), \(h\left(\chi \right) = \sqrt{1-
\tilde{\beta}\left\vert E\right\vert^\upsilon}\); \(\upsilon=1,2\) with \(\tilde{\beta}=\beta /E_{p}\).

\subsection{\mdseries{Rainbow functions pair \(f\left(\chi \right) =\left( 1-
\tilde{\beta}\left\vert E\right\vert \right) ^{-1}\), \(h\left( 
\chi \right) =1\)}}\label{sec:3:1}

The substitution of the pair of rainbow functions \(f\left( \chi \right) = \left( 1 - \tilde{\beta}\left\vert E \right\vert \right)^{-1}\) and \(h\left( \chi \right) = 1\) in the result (\ref{III.13}) yields
\begin{equation}
E^{2} = \mathcal{\tilde{G}}_{nm}\left( 1 - \tilde{\beta}\left\vert E \right\vert \right)^{2};\; \mathcal{\tilde{G}}_{nm} = \mathcal{G}_{nm} + M^{2}.
\label{III.1.1}
\end{equation}
One should notice that \(\left\vert E \right\vert = E_{+}\) and \(\left\vert E \right\vert = -E_{-}\) for the test particles' and antiparticles', respectively. Therefore, it would result in
\begin{equation}
E_{+}^{2}\left( 1 - \tilde{\beta}^{2} \mathcal{\tilde{G}}_{nm}\right) + 2 \tilde{\beta} \, \mathcal{\tilde{G}}_{nm} E_{+} - \mathcal{\tilde{G}}_{nm} = 0,
\label{III.1.2}
\end{equation}
to imply
\begin{equation}
E_{+} = \frac{-\tilde{\beta} \, \mathcal{\tilde{G}}_{nm} + \sqrt{\mathcal{\tilde{G}}_{nm}}}{\left( 1 - \tilde{\beta}^{2} \mathcal{\tilde{G}}_{nm} \right)} \Longrightarrow E_{+} = \frac{\sqrt{\mathcal{\tilde{G}}_{nm}}}{1 + \tilde{\beta} \sqrt{\mathcal{\tilde{G}}_{nm}}}
\label{III.1.3}
\end{equation}
for particles, and
\begin{equation}
E_{-}^{2} \left( 1 - \tilde{\beta}^{2} \mathcal{\tilde{G}}_{nm} \right) - 2 \tilde{\beta} \, \mathcal{\tilde{G}}_{nm} E_{-} - \mathcal{\tilde{G}}_{nm} = 0,
\label{III.1.4}
\end{equation}
to yield
\begin{equation}
E_{-} = \frac{\tilde{\beta} \, \mathcal{\tilde{G}}_{nm} - \sqrt{\mathcal{\tilde{G}}_{nm}}}{\left( 1 - \tilde{\beta}^{2} \mathcal{\tilde{G}}_{nm} \right)} = -\frac{\sqrt{\mathcal{\tilde{G}}_{nm}}}{1 + \tilde{\beta} \sqrt{\mathcal{\tilde{G}}_{nm}}}
\label{III.1.5}
\end{equation}
for the antiparticles. One may very well cast the corresponding energies as
\begin{equation}
E_{nm} = \pm \frac{\sqrt{\mathcal{\tilde{G}}_{nm}}}{1 + \tilde{\beta} \sqrt{\mathcal{\tilde{G}}_{nm}}}
\label{III.1.6}
\end{equation}
It is obvious that the energies for the KG particles and antiparticles are symmetric with respect to the value \(E = 0\). Yet, it is interesting to note that the asymptotic convergence tendency of the values of \(\left\vert E_{\pm} \right\vert\) as \(\Lambda \gg 1\) is \(\left\vert E_{\pm} \right\vert \sim 2 = 1/\tilde{\beta} = E_{p}/\beta\). This would, in turn, result in \(\left\vert E_{\pm} \right\vert \leq E_{p} \Longrightarrow \left\vert E_{\pm} \right\vert_{\max} = E_{p}\) (for \(\beta_{\min} = 1\)). This is documented in Figure \ref{fig2}.

\subsection{\mdseries{Rainbow functions pair $f\left(\chi \right) =\left( 1-
\tilde{\beta}\left\vert E\right\vert \right) ^{-1}=h\left( 
\chi \right) $}}\label{sec:3:2}

\begin{figure*}[ht!]
\centering
\includegraphics[width=0.25\textwidth]{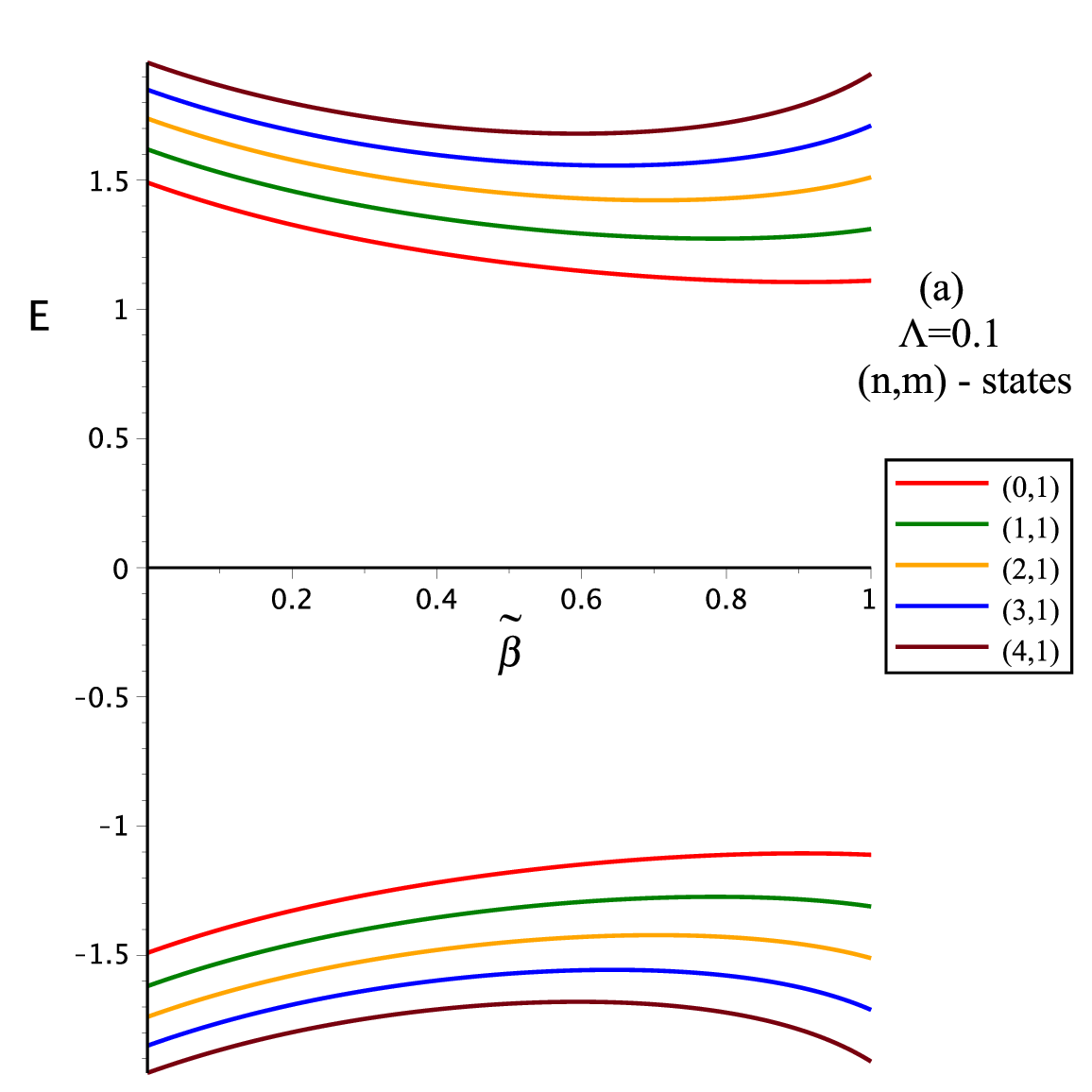}
\includegraphics[width=0.25\textwidth]{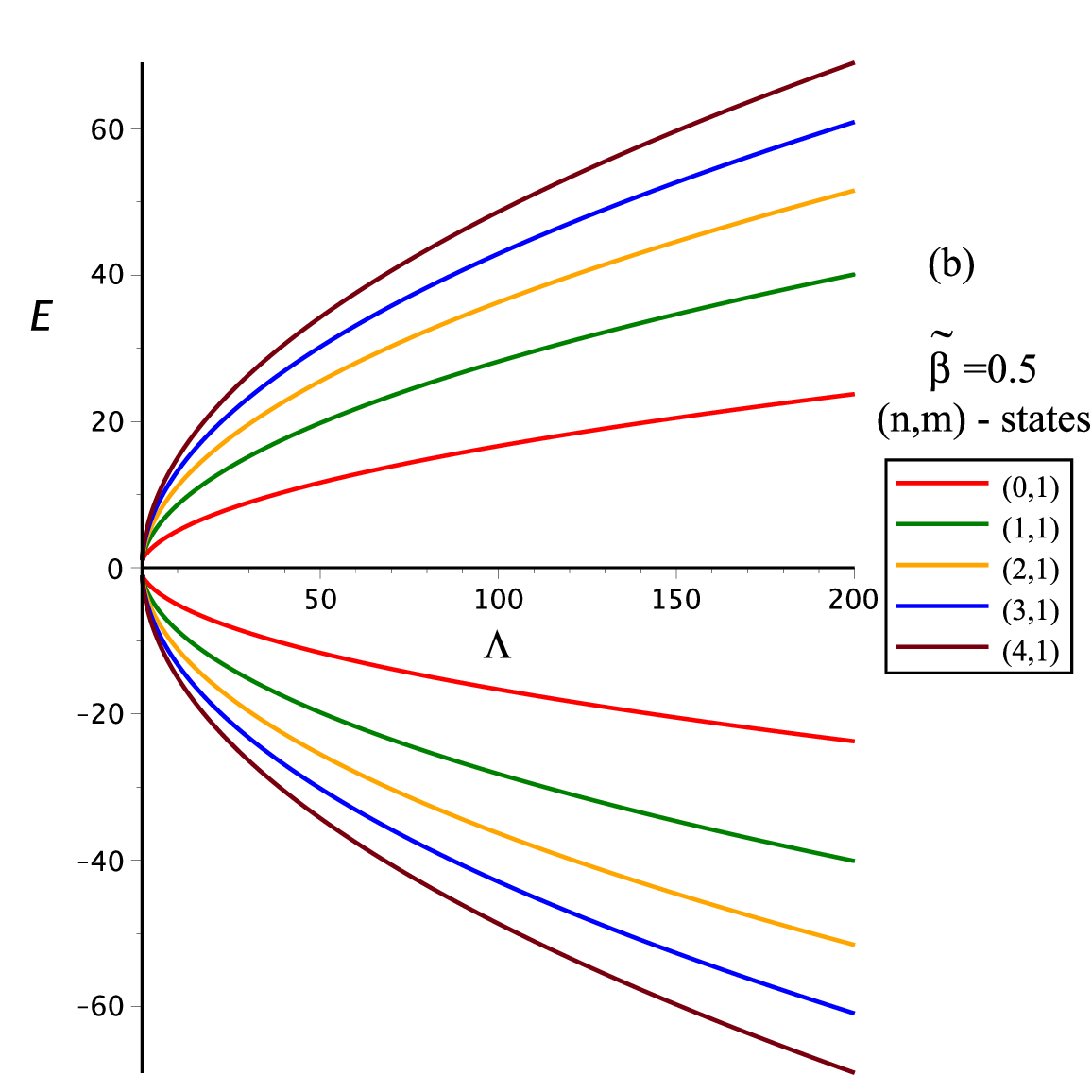}
\includegraphics[width=0.25\textwidth]{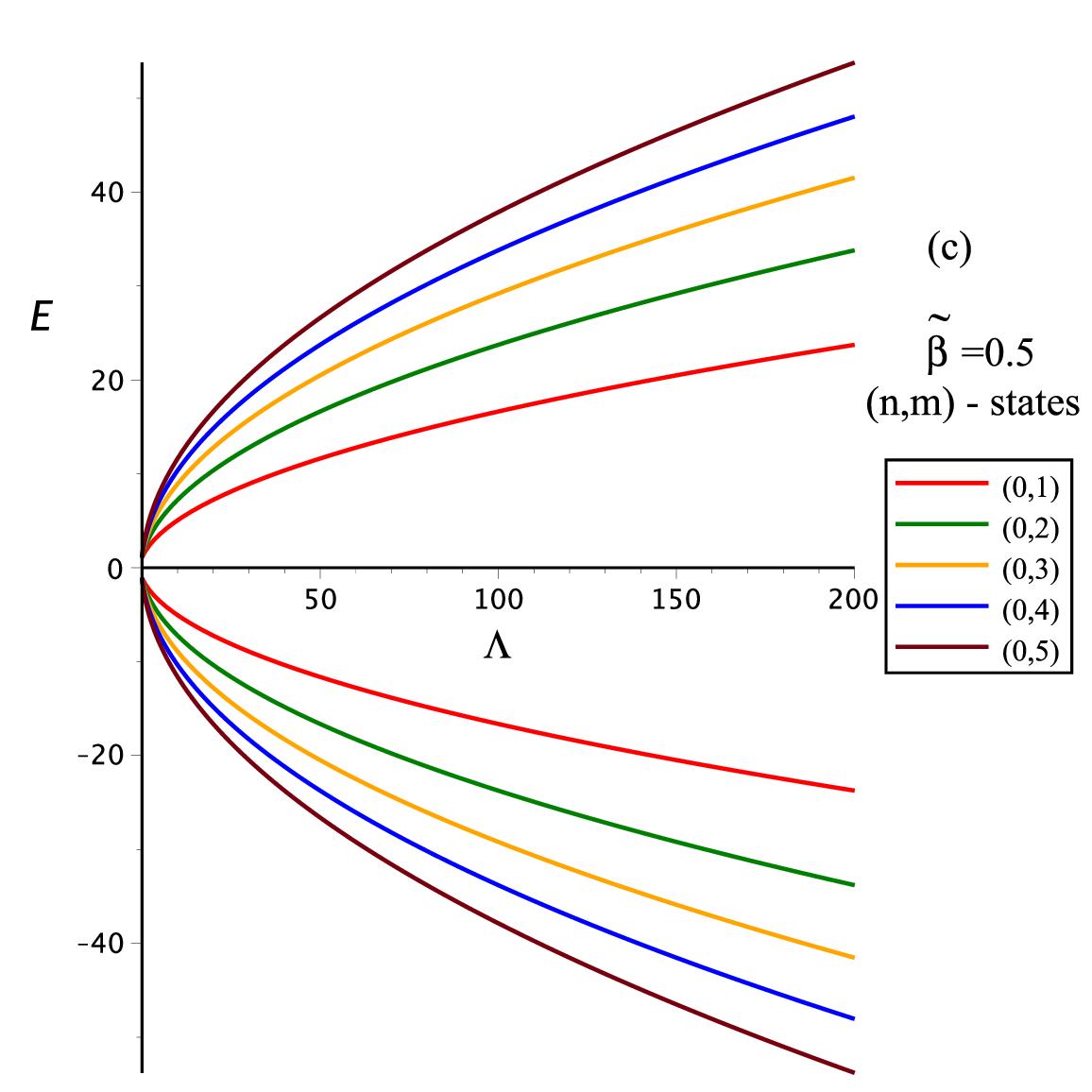}
\caption{\footnotesize  The figure shows the energy levels for KG particles and antiparticles given by (\ref{III.2.1}), where we plot (a) \(E\) against \(\tilde{\beta}\) for \(m=1\) and \(n=0,1,2,3,4\), (b) \(E\) against the cosmological constant \(\Lambda\) for \(m=1\), \(n=0,1,2,3,4\), and \(\tilde{\beta}=0.5\), and (c) \(E\) against the cosmological constant \(\Lambda\) for \(n=0\), \(m=1,2,3,4,5\), and \(\tilde{\beta}=0.5\).}
\label{fig3}
\end{figure*}

Upon the substitution of such rainbow functions pair in the result (\ref{III.13}) on obtains%
\begin{equation}
    E^2(1-\tilde{\beta}^2M^2)+2M^2\tilde{\beta}|E|-\mathcal{\tilde{G}}_{nm}=0. \label{III.2.1}
\end{equation}%
We now follow the same recipe as above and report%
\begin{equation}
    E_+=\frac{-\tilde{\beta}M^2+\sqrt{\tilde{\beta}^2M^2+\mathcal{\tilde{G}}_{nm}(1-\tilde{\beta}^2M^2)}}{(1-\tilde{\beta}^2M^2)}, \label{III.2.2}
\end{equation}%
and%
\begin{equation}
    E_-=\frac{\tilde{\beta}M^2-\sqrt{\tilde{\beta}^2M^2+\mathcal{\tilde{G}}_{nm}(1-\tilde{\beta}^2M^2)}}{(1-\tilde{\beta}^2M^2)}.  \label{III.2.3}
\end{equation}%
The symmetrization of the energy levels with respect to the $E=0$ value is obvious.
\begin{figure*}[ht!]
\centering
\includegraphics[width=0.25\textwidth]{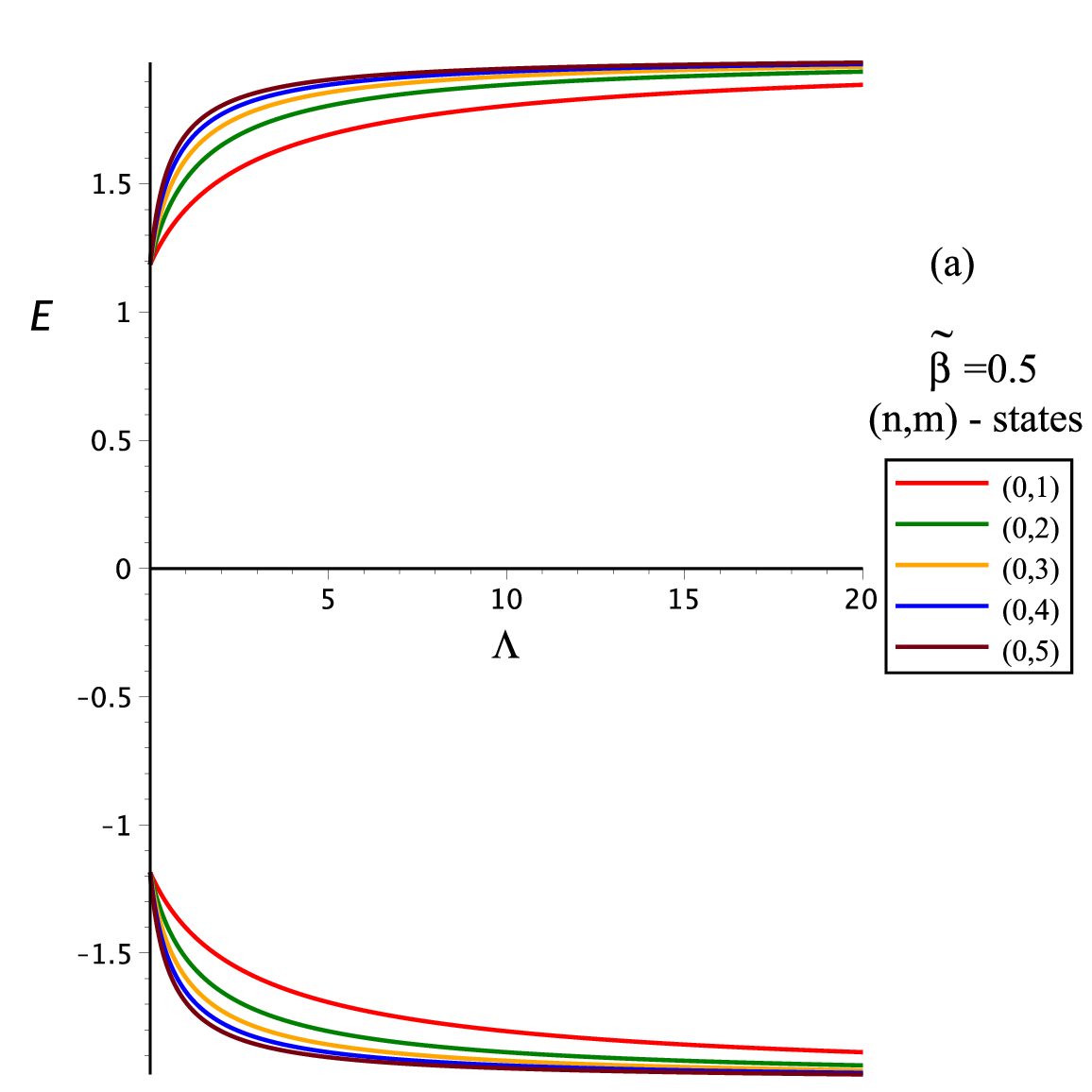}
\includegraphics[width=0.25\textwidth]{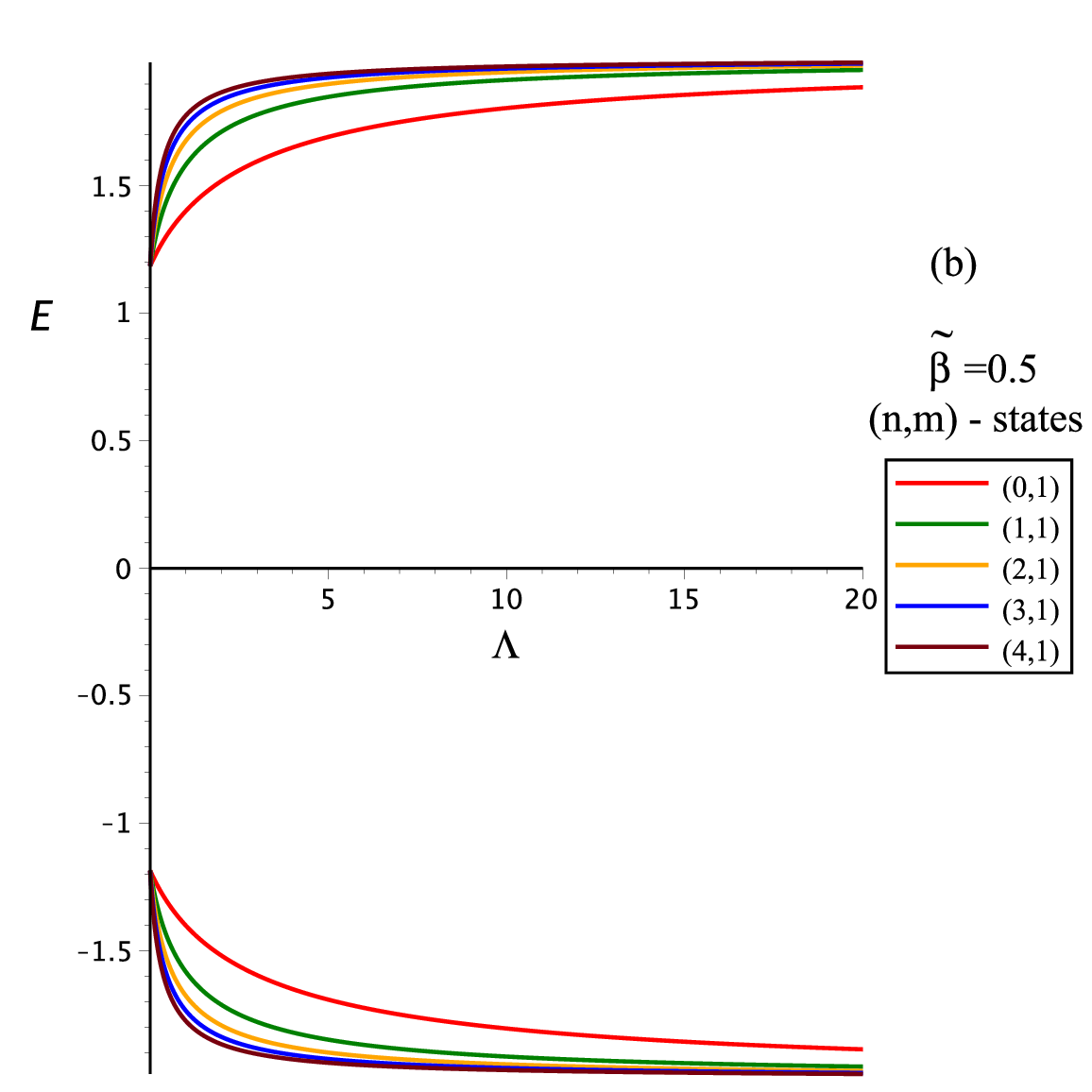}
\caption{\footnotesize  The figure shows the energy levels for KG particles and antiparticles given by (\ref{III.3.1}), so that we plot (a) $E$ against $\tilde{\beta}$ for $m=1$ and $n=0,1,2,3,4$, (b) $E$ against the cosmological constant $\Lambda$ for $m=1$, $n=0,1,2,3,4$ and $\tilde{\beta}=0.5$, and (c) $E$ against the cosmological constant $\Lambda$ for $n=0$, $m=1,2,3,4,5$ and $\tilde{\beta}=0.5$.}
\label{fig4}
\end{figure*}
\begin{figure*}[ht!]
\centering
\includegraphics[width=0.25\textwidth]{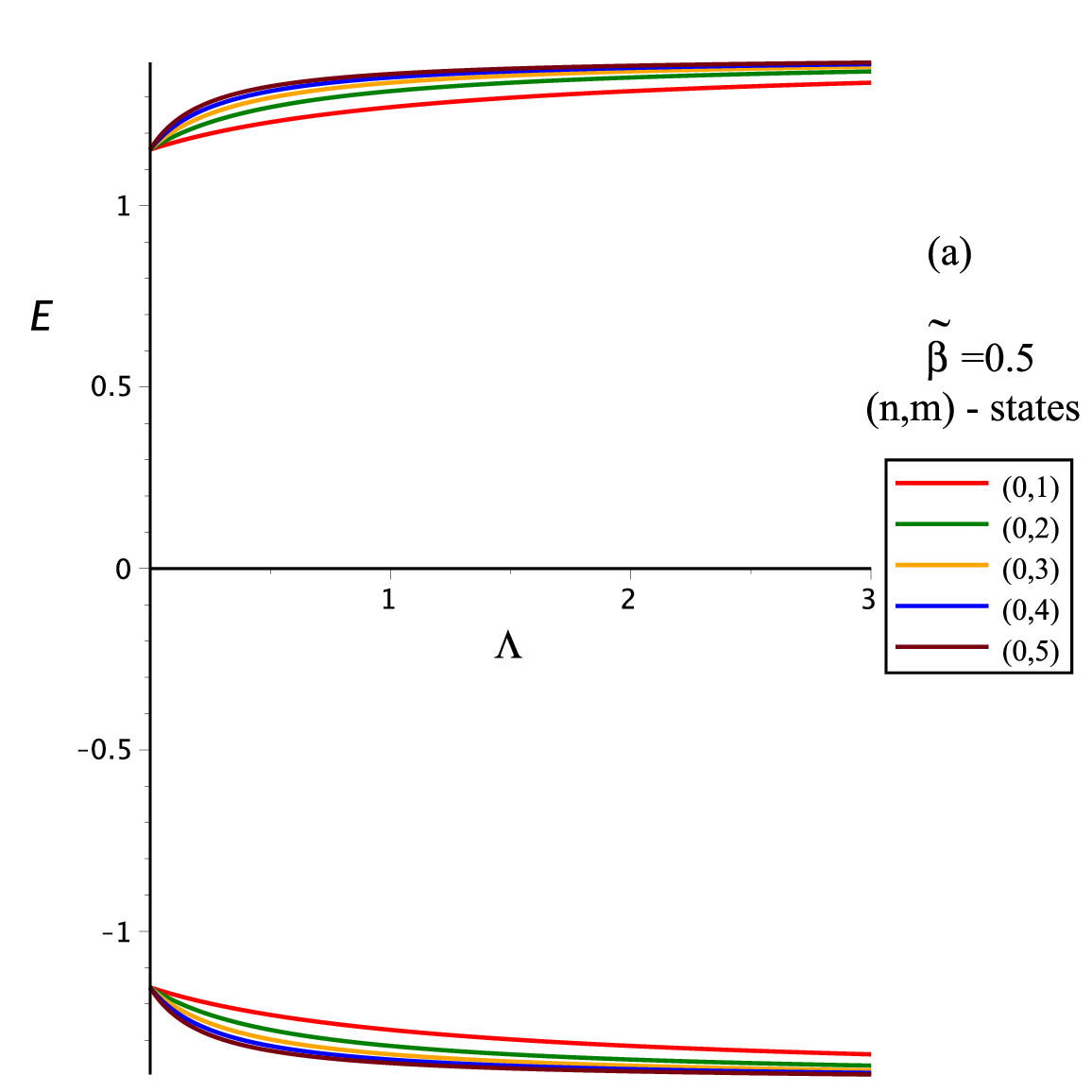}
\includegraphics[width=0.25\textwidth]{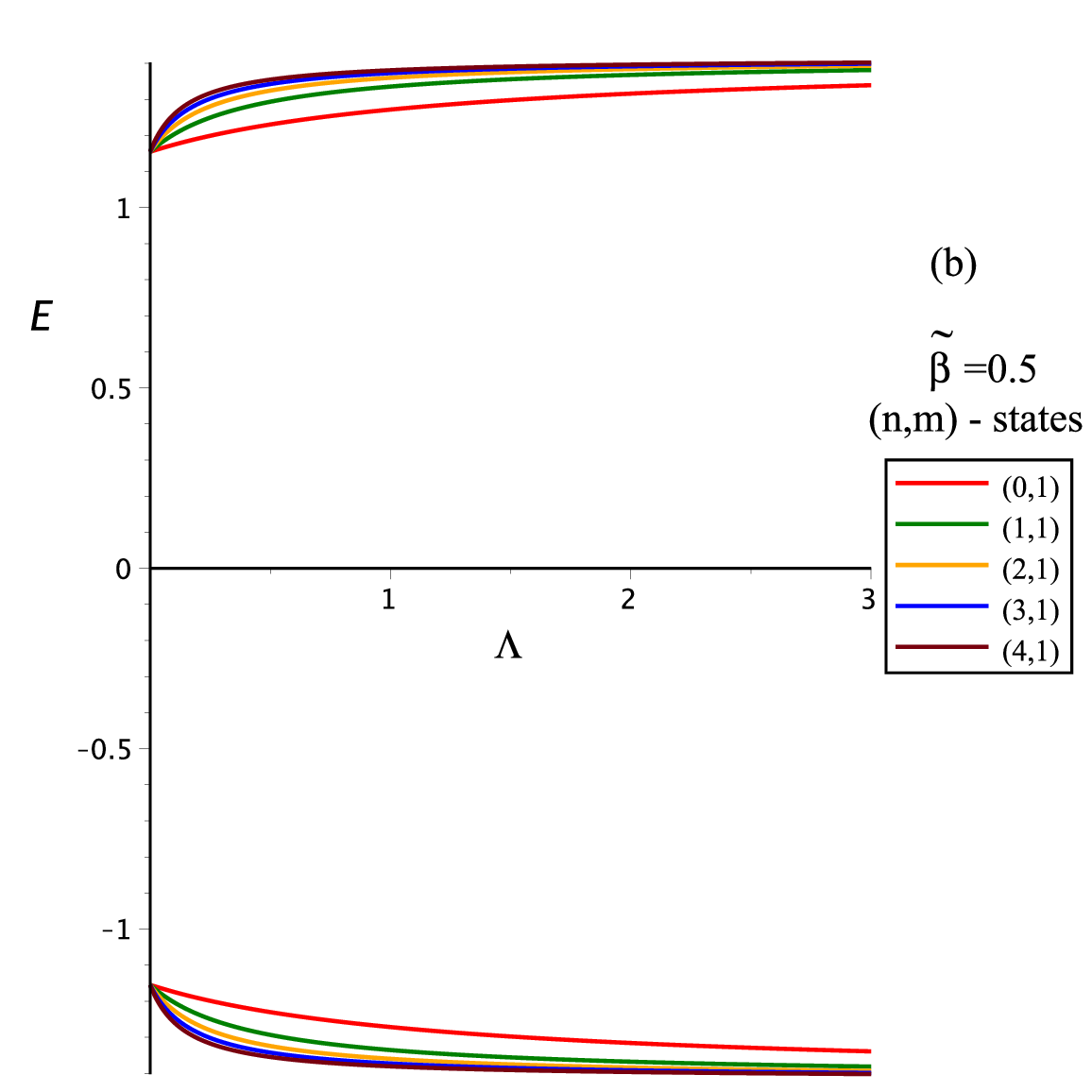}
\caption{\footnotesize  The figure shows the energy levels for KG particles and antiparticles given by (\ref{III.3.4}), so that we plot (a) $E$ against $\tilde{\beta}$ for $m=1$ and $n=0,1,2,3,4$, (b) $E$ against the cosmological constant $\Lambda$ for $m=1$, $n=0,1,2,3,4$ and $\tilde{\beta}=0.5$, and (c) $E$ against the cosmological constant $\Lambda$ for $n=0$, $m=1,2,3,4,5$ and $\tilde{\beta}=0.5$.}
\label{fig5}
\end{figure*}
In Figure \ref{fig3}, we observe that there is no tendency for the energies to converge to \(|E|_{max}=E_p\), but, on the contrary, the energies grow with a growing cosmological constant \(\Lambda\). Here, in addition to the fact that this pair of rainbow gravity functions may solve the horizon problem but does not produce a varying $c$ (as mentioned in \cite{R1.24}), it does not comply with the rainbow gravity theory that \(|E|\leq E_p\). 

\subsection{\mdseries{Rainbow functions pair $f(\chi)=1$, $h\left(\chi \right) = \sqrt{1-
\tilde{\beta}(\left\vert E\right\vert)^\upsilon}$\,; $\upsilon=1,2$ }}\label{sec:3:3}

Such rainbow function pairs are loop quantum gravity-motivated pairs \cite{R1.38,R1.39} that are found to completely comply with rainbow gravity since they ensure that the Planck energy $E_p$ is the maximum possible energy for particles and antiparticles (e.g., \cite{R1.34,R1.35,R1.40}). It is therefore interesting to investigate their performance on KG particles/antiparticles in magnetized BM-spacetime. We start with $\upsilon=1,$ and the result in (\ref{III.13}) would imply that
\begin{equation}
    E^2+\tilde{\beta}\,\mathcal{G}_{nm}\,|E|-\tilde{\mathcal{G}}_{nm}=0. \label{III.3.1}
\end{equation}
In this case, one obtains
\begin{figure*}[ht!]
\centering
\includegraphics[width=0.25\textwidth]{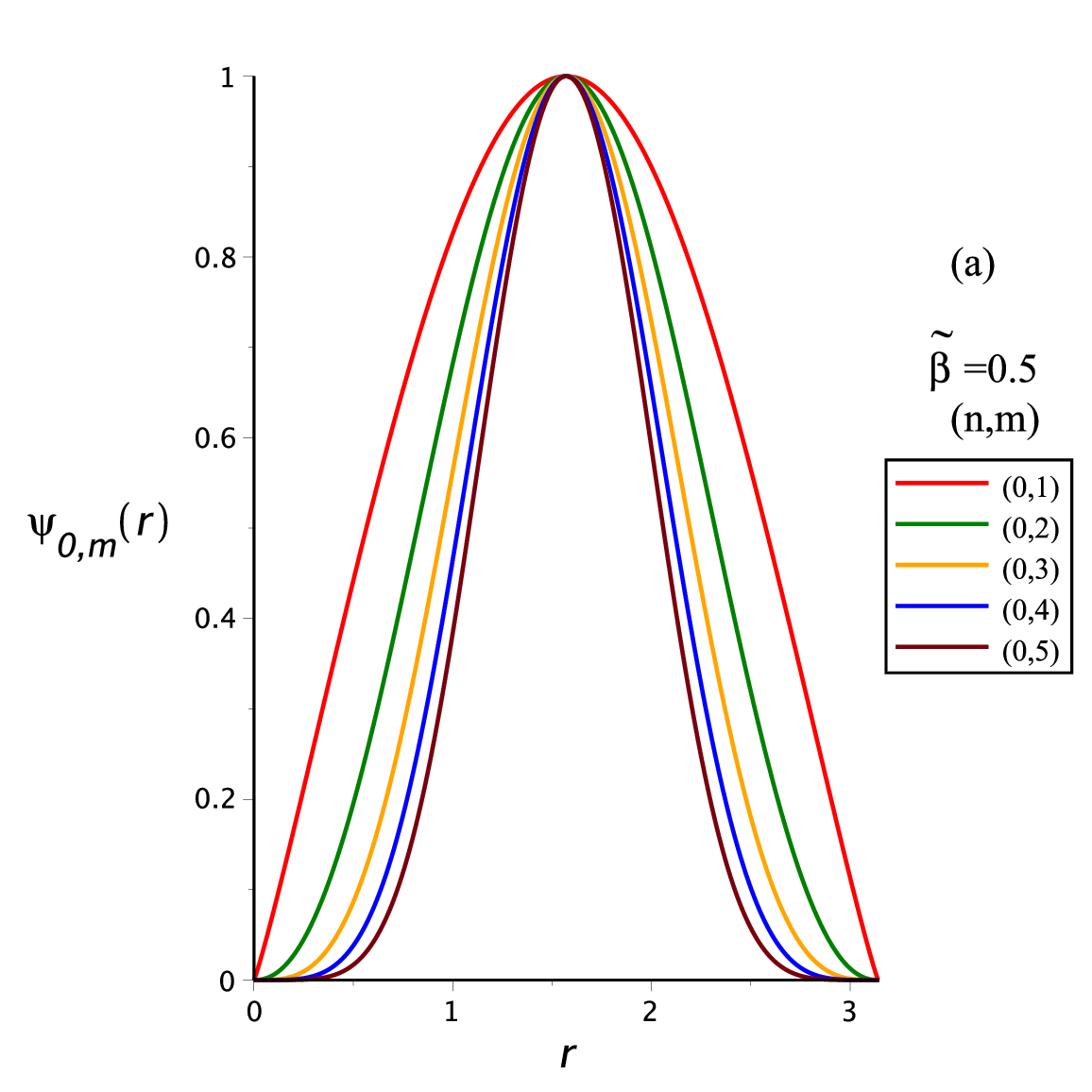}
\includegraphics[width=0.25\textwidth]{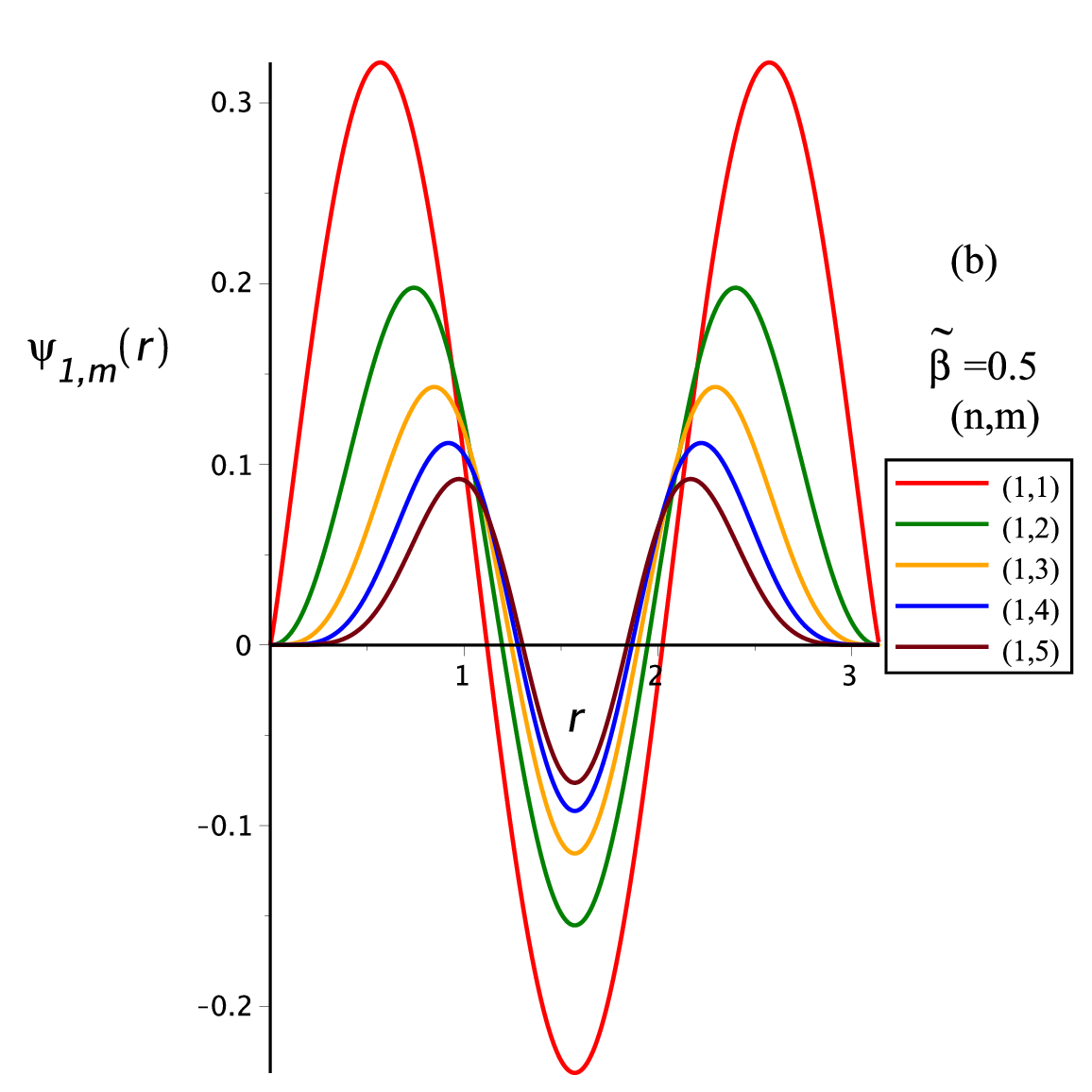}
\includegraphics[width=0.25\textwidth]{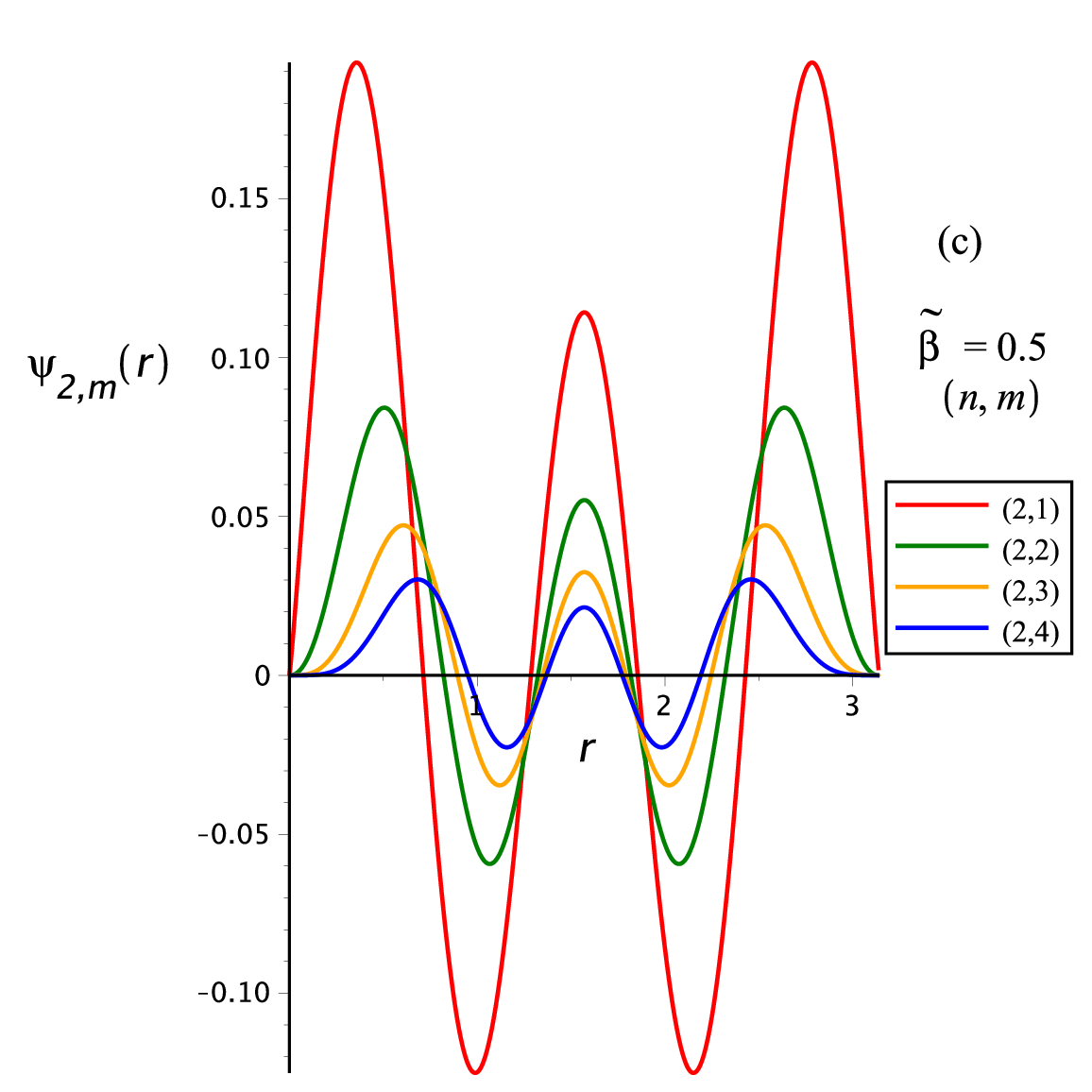}
\caption{\footnotesize  The figure shows the radial wave functions $\psi(r)$ in (\ref{III.12}), for $\alpha=0.9,\,\tilde{\beta}=0.5,\, M=1=k$ and $m=1,2,3,4$ we plot $\psi(r)$ so that (a) for $n=0$, (b) for $n=1$, and (c) for $n=2$.}
\label{fig6}
\end{figure*}
\begin{equation}
    E_+=\frac{-\tilde{\beta}\,\mathcal{G}_{nm}}{2}+\frac{1}{2}\sqrt{\tilde{\beta}^2\,\mathcal{G}_{nm}^2+4\tilde{\mathcal{G}}_{nm}} \label{III.3.2}
\end{equation}
and
\begin{equation}
 E_-=\frac{\tilde{\beta}\mathcal{G}_{nm}}{2}-\frac{1}{2}\sqrt{\tilde{\beta}^2\mathcal{G}_{nm}^2+4\tilde{\mathcal{G}}_{nm}} \label{III.3.3}
\end{equation}%
These energy levels are plotted in Figure \ref{fig4}. We observe that as the cosmological constant $\Lambda\rightarrow\infty$, our result in (\ref{III.3.1}) would imply that $|E|_{max}\sim1/\tilde{\beta}=2$. This is clearly the tendency of the asymptotic convergence of the KG particles/antiparticles energies in BM-spacetime, as observed in Figure \ref{fig4}. We now consider $\upsilon=2$ to obtain
\begin{equation}
    E^2=\frac{\tilde{\mathcal{G}}_{nm}}{1+\tilde{\beta}\,\mathcal{G}_{nm}}\Rightarrow E_\pm=\pm\sqrt{\frac{\tilde{\mathcal{G}}_{nm}}{1+\tilde{\beta}\,\mathcal{G}_{nm}}}. \label{III.3.4}
\end{equation}
The corresponding energy levels are plotted in Figure \ref{fig5}, which in turn documents the asymptotic tendency of the energy levels as the cosmological constant $\Lambda\rightarrow\infty \Rightarrow|E|_{max}\sim \sqrt{1/\tilde{\beta}}=\sqrt{2}\approx 1.41$ (as mandated by (\ref{III.3.4})).

\section{\mdseries{Concluding remarks and open problems }}\label{sec:4}

In this work, we have investigated the effect of the gravitational field produced by Bonner-Melvin spacetime, with a cosmological constant, in rainbow gravity on the Klein-Gordon bosons. We have shown that such a particular spacetime manifestly and unavoidably introduces domain walls at \(r=0,\pi\). This, in turn, would not only agree with \v{Z}ofka's \cite{R1.1} proposal that $\sqrt{2\Lambda }r=\pi$ (i.e., with our $r_{our}=\sqrt{2\Lambda }r$ of \v{Z}ofka) as the location of an axis of some sort, but it also very well identifies it as a domain wall (one of the known topological defects). In fact, the very existence of the sinusoidal term, \(\sin(\sqrt{2\Lambda}r)^2\), in the BM-spacetime metric (\ref{I.3}) suggests that \(\sin(\sqrt{2\Lambda}r)^2\in[0,1]\) introduces the singularities associated with \(r=0, \pi\) and consequently and effectively generates two domain walls, at \(r=0, \pi\), that confine KG particles/antiparticles to move within.

We have shown, with brute force evidence, that the commonly used approximation assumption that \(r<<1\rightarrow \sin(r)\sim r\Rightarrow \tan(r)\sim r\) (e.g. \cite{R1.4,R1.4.1}) is not valid because it would change not only the dynamics of the quantum mechanical system at hand but also eliminate the domain walls, the very characterization of the magnetized BM spacetime (mentioned above and documented in Figures \ref{fig1}(a) and \ref{fig1}(b)).

Under such BM-spacetime domain wall settings, we have discussed the effects of rainbow gravity on KG particles/antiparticles using the rainbow function pairs (i) \(f\left( \chi \right) =1/\left( 1-\tilde{\beta}\left\vert E\right\vert \right),\,h(\chi)=1\), (ii) \(f(\chi)=\left(1-\tilde{\beta}\left\vert E\right\vert \right) ^{-1}=f(\chi)\), and (iii) $f(\chi)=1$, $h\left(\chi \right) = \sqrt{1-%
\tilde{\beta}(\left\vert E\right\vert)^\upsilon}$\,; $\upsilon=1,2$ . The rainbow function pairs in (i) and (iii) are found to fully comply with the rainbow gravity theory for they ensure that the Planck energy $E_p$ is the maximum possible energy for particles and antiparticles (e.g., \cite{R1.34,R1.35,R1.40}). However, the pair of rainbow functions in (ii) is found to have no response to rainbow gravity at all for massless KG particles and antiparticles and a mild response to rainbow gravity for KG particles and antiparticles with \(M\neq 0\).

In connection with the radial wave functions $\psi_{n,m}(r)$ in (\ref{III.12}), moreover, it is convenient and interesting to report, with \(A_0=1\), that
\begin{gather}
 \psi_{0,m}(r) =\mathcal{C}\, \sin(r)^{|\tilde{m}|};\quad \text{for}\,\, n=0, \label{IV.1} \\
 \psi_{1,m}(r) =\mathcal{C}\,\sin(r)^{|\tilde{m}|} \left[1+A_1\,\sin(r)^2\right];\quad \text{for}\,\, n=1, \label{IV.2} \\
\psi_{2,m}(r) =\mathcal{C}\,\sin(r)^{|\tilde{m}|} \left[1+A_1\,\sin(r)^2+A_2\,\sin(r)^4\right];\quad \text{for}\,\, n=2. \label{IV.3}
\end{gather}
Where for each value of $n$ one should use the correlation (\ref{III.91}) so that
\begin{gather}
    A_1=-\frac{2|\tilde{m}|+3}{2|\tilde{m}|+2}, \text{for}\, n=1,\,\text{and} \label{IV.4}\\
    A_1=-\frac{2|\tilde{m}|+5}{|\tilde{m}|+1},\, A_2= \frac{4\tilde{m}^2+24|\tilde{m}|+35}{4(|\tilde{m}|+1)(|\tilde{m}|+2)}, \text{for}\, n=2. \label{IV.5}
\end{gather}
A sample of the first three radial wave functions is plotted in Figure \ref{fig6}. The general behavior of which fits very well the textbook one. However, we should be aware that our hypergeometric polynomial of order $n$ in (\ref{III.12}) is of even powers of \(\sin(r)\). This would immediately suggest the correlation between the radial quantum number $n_r$ (i.e., the number of nodes in the radial wave function) and the order $n$ of the hypergeometric polynomial to read $n_r=2n$. This correlation is very clear in Figure \ref{fig6}, where one observes no nodes for $n=0=n_r$ (see Fig. \ref{fig6}(a)), two nodes for $n=1\Rightarrow n_r=2$ (see Fig. \ref{fig6}(b)), and four nodes for $n=2\Rightarrow n_r=4$ (see Fig. \ref{fig6}(c)). Our analysis reveals that, while domain walls constrain the radial motion, both the angular motion and the linear motion along the \(z\) direction remain permissible, regardless of whether the particles are massless or massive. This suggests that the magnetic background can enable the formation of rotating matter rings (with \( |\tilde{m}| \neq 0 \)), which may be static with \(k = 0\) or in motion with \(k \neq 0\), across both extremely high-energy and low-energy regimes. This phenomenon arises from the fact that the effects of rainbow gravity do not affect the radially allowed region, in which the test fields are confined, despite potentially influencing the symmetry breaking between particle and antiparticle energy levels near zero energy.

\vspace{0.15cm}
\setlength{\parindent}{0pt}

\begin{figure}[ht!]
\centering
\includegraphics[width=0.4\textwidth]{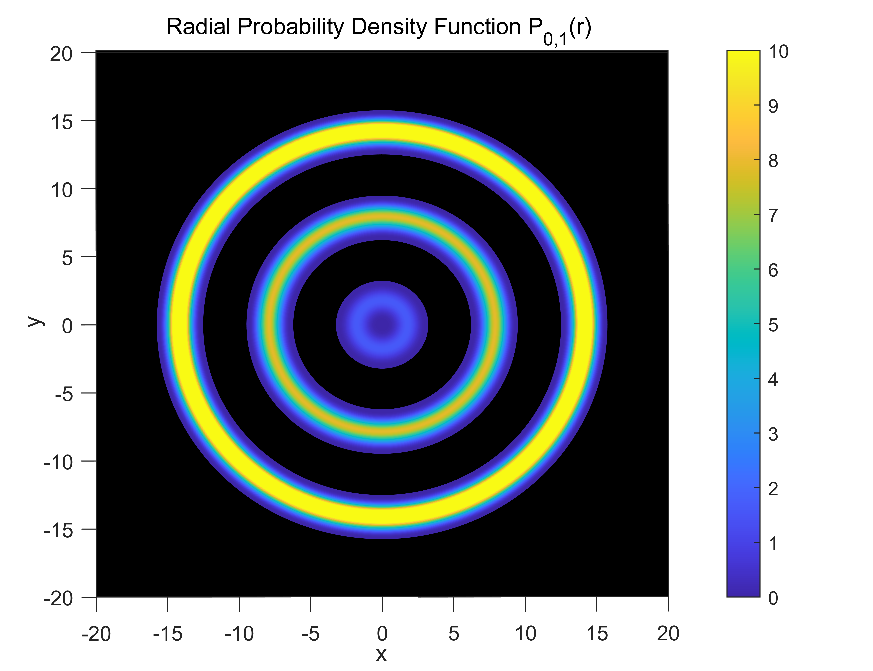}
\caption{\footnotesize This plot shows the radial probability density for the bosonic ground state with \(n=0\) and \(m=1\) (noting that \(|m| \neq 0\)), as a function of position in two dimensions. For simplicity, we set \(\mathcal{C} = 1\).}
\label{fig7}
\end{figure}

On the other hand, magnetic domain walls are intrinsic features of condensed matter systems, emerging from spatial variations in magnetization within ferromagnetic or antiferromagnetic materials. These domain walls are crucial for understanding phenomena such as magnetoresistance, spintronics, and topological effects in low-dimensional systems. In some condensed matter systems, the dynamics of these domain walls can exhibit similarities to cosmological magnetic structures, where field configurations display topologically nontrivial behavior. Models of the cosmological magnetic universe, particularly those describing large-scale structures such as cosmic domain walls, provide a conceptual framework for exploring the dynamics of domain walls in condensed matter systems. The resemblance between the behavior of domain walls in condensed matter systems and those in cosmological contexts suggests that cosmological models, such as the \(2+1\)-dimensional magnetic BM spacetime, can be adapted to investigate the underlying physics of magnetic domain walls, offering insights into their formation and evolution in material systems under extreme conditions. The \(2+1\)-dimensional magnetic BM spacetime (without point-like defect) is given by \cite{AO-2}:  
\[
ds^{2} = -c^2dt^{2} + dr^{2} + \alpha^{2}\sin ^{2}\left( \sqrt{2\Lambda} r\right)\, d\varphi^{2}. \tag{C.1}
\]
This model can aid in understanding the dynamics of magnetic domain walls under various physical conditions. The effect of an out-of-plane magnetic field in this spacetime background can be described through the angular component of the electromagnetic vector potential \cite{AAF,castro}:  
\[
A_{\varphi} = \alpha \, \sin \left( \sqrt{2\, \Lambda}\, r \right) f(r)\approx \alpha \, \sqrt{2 \,\Lambda} \, r \, f(r) \quad \text{for} \quad \Lambda \ll 1.
\]
As \(\Lambda \to 0\), the Gaussian curvature of the spacetime, \(K = 2\Lambda\), tends to zero except at \(r = 0\). In this regime, the system may describe conditions where either the magnetic field is weak or the spacetime curvature is nearly vanishing, which holds when \(\Lambda \ll 1\). This approximation can be particularly useful for understanding the dynamics of charge carriers or photonic modes in weak magnetic fields, potentially explaining discrepancies between theoretical predictions and experimental results that may arise due to weak magnetic fields or a minimal baseline potential (nonzero). Consequently, this approach could be valuable for modeling the behavior of charge carriers and electromagnetic waves under various environmental conditions. Under the weak-field approximation, the magnetic field primarily influences the total angular momentum terms. This suggests that the effective electromagnetic potential can be approximated as \(\mathcal{A}_{\varphi} \sim \frac{1}{\alpha \sqrt{2 \Lambda} r}\), which remains significant at \(r = 0\). In this context, the field-angular momentum interactions bear a resemblance to those of a rotating magnetic vortex. A magnetic vortex is a topological defect in which magnetic field lines form a swirling structure around a central point, preventing the magnetization from smoothly transitioning into a uniform state. This structure provides the vortex with unique topological stability, which is particularly relevant for applications in information storage and spintronics. The core of the vortex typically exhibits a distinct magnetization direction, contrasting with the surrounding spins that align in a rotating pattern. These vortices are critical for nanotechnology and spintronic devices, as they play an essential role in various magnetic phenomena and applications. However, this effect vanishes when considering zero angular momentum states (S-states). In our case, without rainbow gravity effects, the regular solution function for $n=0$ and $m=1$ (note that $|m|\neq 0$) is:
\begin{equation}
U_{0,1}=\mathcal{C}\,\sin(r)^{3/2}. \label{SF}   
\end{equation}
The radial probability density function is given by \( P_{0,1}(r) = \int_{0}^{r} |U_{0,1}|^2 r \, dr \), which simplifies to \( P_{0,1}(r) = \int_{0}^{r} |\mathcal{C}|^2 \sin(r)^3 r \, dr \). By introducing the transformations \( x = r \cos{\varphi} \) and \( y = r \sin{\varphi} \), we visualize the radial probability density function as a function of position in two dimensions, as shown in Figure \ref{fig7}. This figure reveals that the modes are restricted to rotating, ring-like structures (note that \( |m| \neq 0 \)), or, in other words, the bosonic states manifest as spinning vortices. It is evident that, in principle, one can influence the behavior of these spinning vortices by adjusting the out-of-plane magnetic field. This is further demonstrated by rewriting the solution function as \( U(r) \propto \sin(\sqrt{2\Lambda} r) \), which indicates that the corresponding states generate magnetized spinning vortices in monolayer materials.
Moreover, the \(2+1\)-dimensional spacetime under consideration is characterized by a constant positive Gaussian curvature. Accordingly, the corresponding surface can be effectively modeled as having a space- and frequency-dependent refractive index, \(n\). Depending on the effective curvature, the refractive index can be negative, positive, or purely imaginary for different frequency regimes, such as visible light frequencies or the \(\gamma\)-ray regime. In principle, this suggests that the out-of-plane aligned magnetic field can effectively modulate the refractive index as needed, thereby influencing photonic modes and the evolution of quasi-particles in low-dimensional materials (see also \cite{new}).

\section*{\small{Data availability}}
This is a theoretical research study, including equations and derived quantities presented in the main text of the paper.

\section*{\small{Conflicts of interest statement}}
The authors have disclosed no conflicts of interest.

\section*{\small{Funding}}
This research has not received any funding.

\end{document}